\documentclass[conference,compsoc]{IEEEtran}

\ifCLASSOPTIONcompsoc
  \usepackage[nocompress]{cite}
\else
  \usepackage{cite}
\fi

\ifCLASSINFOpdf
\else
\fi

\usepackage{graphicx}
\usepackage[dvipsnames]{xcolor}
\usepackage{booktabs}
\usepackage{footmisc}
\usepackage{hyperref}

\newcommand{\THtool}{the TH tool}

\newcommand{\THmodel}{\textit{threat hunter process of building a mental model}}

\newcommand{\Enrich}{\textit{Enrich}}

\usepackage{fancyhdr}
\usepackage{lastpage}

\fancypagestyle{firststyle}{
   \fancyhf{}
    \fancyfoot[L]{{\footnotesize Author's version. Accepted for publication at Computers \& Security Journal (ISSN: 0167-4048)}}
    \fancyfoot[R]{\small Page \thepage \hspace{1pt} of \pageref{LastPage}}
}

\begin{document}

\title{Fuzzy to Clear: \\ Elucidating the Threat Hunter Cognitive Process and Cognitive Support Needs}

\author{\IEEEauthorblockN{
Alessandra Maciel Paz Milani,\IEEEauthorrefmark{1}
Arty Starr,\IEEEauthorrefmark{1}
Samantha Hill,\IEEEauthorrefmark{1} 
Callum Curtis,\IEEEauthorrefmark{1} \\
Norman Anderson,\IEEEauthorrefmark{1} 
David Moreno-Lumbreras\IEEEauthorrefmark{2} and
Margaret-Anne Storey\IEEEauthorrefmark{1}
\IEEEauthorblockA{
\IEEEauthorrefmark{1}University of Victoria, Victoria, Canada\\
\IEEEauthorrefmark{2}Universidad Rey Juan Carlos, Madrid, Spain}}
 amilani@uvic.ca,
 artystarr@uvic.ca,
 samymhill@gmail.com,
 callumcurtis@uvic.ca, \\
 normananderson@uvic.ca, 
 david.morenolu@urjc.es and
 mstorey@uvic.ca}

\maketitle

\thispagestyle{firststyle}

\begin{abstract}
With security threats increasing in frequency and severity, it is critical that we consider the important role of threat hunters. These highly-trained security professionals learn to see, identify, and intercept security threats. Many recent works and existing tools in cybersecurity are focused on automating the threat hunting process, often overlooking the critical human element. Our study shifts this paradigm by emphasizing a human-centered approach to understanding the lived experiences of threat hunters. By observing threat hunters during hunting sessions and analyzing the rich insights they provide, we seek to advance the understanding of their cognitive processes and the tool support they need. Through an in-depth observational study of threat hunters, we introduce a model of how they build and refine their mental models during threat hunting sessions. We also present 23 themes that provide a foundation to better understand threat hunter needs and suggest five actionable design propositions to enhance the tools that support them. Through these contributions, our work enriches the theoretical understanding of threat hunting and provides practical insights for designing more effective, human-centered cybersecurity tools.
\end{abstract}


\IEEEpeerreviewmaketitle

\section{Introduction}
\label{sec_introduction}

The complexity and challenges of the cybersecurity landscape have dramatically increased over the past several years. In 2023 alone, 2,365 cyber attacks were reported in the United States, with an estimated 343,338,964 victims,\footnote{\url{https://www.forbes.com/advisor/education/it-and-tech/cybersecurity-statistics/}} while another global threat report indicates the fastest observed breakout time for interactive eCrime intrusion was only 2 minutes and 7 seconds.\footnote{\url{https://www.crowdstrike.com/en-us/global-threat-report/}} 
As the frequency, speed, and severity of cyber threats continue to rise, it is more crucial than ever for organizations to implement robust security measures. One such measure is the practice of threat hunting, which plays a critical role in combating increasingly sophisticated and evolving cyber threats \cite{mahboubi_evolving_2024}.

Threat hunting is the proactive process of searching IT infrastructure for signs of malicious activity and suspicious behaviours that have evaded existing security defences \cite{vanos2018}.
It involves comprehensive data analysis from various sources, including network traffic, system logs, endpoint telemetry, and threat intelligence feeds, to uncover potential threats \cite{mahboubi_evolving_2024}.
Threat hunting can be performed by different cybersecurity roles, such as \textit{security analysts}, who may be part of dedicated teams or have other organizational responsibilities~\cite{badva2024}.
However, a specialized full-time role has emerged: the \textit{threat hunter}. Throughout our paper, when we say threat hunter (TH), we refer to the subset of cybersecurity professionals for whom threat hunting is their primary role.

More broadly, multiple studies describe the challenges faced by these highly skilled cybersecurity professionals, who are facing growing demand for their time and prone to burnout~\cite{nobles2022, nepal2024}. 
Examples of challenges include the cognitive workload required to sift through large amounts of data~\cite{dykstraCyberOperationsStress2018} and the need to develop situational awareness of complex systems~\cite{ofte2023}. 
Additionally, these professionals must navigate these challenges during the already difficulty process of collaborating with their peers, clients and/or stakeholders \cite{mancuso2020}.
Unlike reactive security strategies that rely on automated threat detection systems' alerts to act, THs need to proactively examine extensive and often complex logs and user activities to identify stealthy threats that may be undetectable by conventional methods \cite{nour2023_th_enterprise}. 
This complicates the TH role beyond the challenges outlined above (as applicable to all cybersecurity professionals). 
These challenges require further investigation into threat hunting practices and the development of effective support for human-in-the-loop approaches throughout this process \cite{nour2023_th_enterprise, mahboubi_evolving_2024, badva2024}.

In this complex context, designing strategies and developing innovative tools to support THs becomes crucial. The problem is that designers lack studies that capture THs' cognitive processes or challenges that could aid the design process. 
Although we can find multiple tools, methodologies and models supporting the threat hunting process~\cite{vanos2018, nour2023_th_enterprise, maxam2024, mahboubi_evolving_2024, badva2024}, there is still a lack of studies and empirical evidence that captures the THs' cognitive processes and the challenges they struggle with (similar limitations are also discussed by other authors on human-centered cybersecurity \cite{andrade_cognitive_2019, mancuso2020}).

Therefore, this work explores the lived experiences of threat hunters through an in-depth observational study. Specifically, we aim to address a gap in the literature by:
(a) observing THs as they perform the tasks of a typical workday;
(b) understanding the THs' cognitive processes and workflows for problem-solving; and
(c) understanding the tooling needs of THs.
This paper presents the findings from a comprehensive analysis of data collected during six observation sessions with four experienced THs from a cybersecurity organization. The participants were skilled in using Security Information and Event Management (SIEM) and User Entity Behavior Analytics (UEBA) tools, their primary tools for conducting threat hunting activities.
 
The contributions from our study are:
\begin{enumerate}
    \item Organization of the main findings and insights from the observation sessions into 23 themes. These insightful themes can serve as a foundation (or inspire system requirements) to better understand THs needs and help designers build better supportive tools; 
    \item A visual representation of the TH's process of building a mental model during a threat hunting session, and 
    \item Proposal of five actionable design propositions to help designers enhance the tools that support THs.    
\end{enumerate}

Ultimately, we hope these contributions serve as a valuable resource to tool designers, cybersecurity practitioners, and the research community as our work fosters a deeper understanding of the \THmodel~and facilitates advancements of new strategies and solutions to support THs.

In this paper, we start by presenting an overview of the background and related work (Section~\ref{sec_related_work}), followed by a description of the study design (Section~\ref{sec_study_design}). 
Next, we present our main findings, which include:
the thematic data analysis results (Section~\ref{sec_results}), 
followed by the introduction of the \THmodel~(Section~\ref{sec_sub_results_model}). 
After that, we summarize the proposed five design propositions (Section~\ref{sec_design_propositions}). Then, we discuss the implications of our proposed model, as well as limitations and future work (Section~\ref{sec_discussion}). 
To conclude, we share our final considerations for the study (Section~\ref{sec_conclusions}).

\section{Background and Related Work}
\label{sec_related_work}

In this paper, we use the term \textit{threat hunter} to refer to an \textit{emerging full-time role}, while we still acknowledge that \textit{threat hunting} is a process or activity that can be performed by various cybersecurity professionals, depending on an organization's structure and maturity level.
For us, the term \textit{threat hunter} distinguishes the individual from the process itself, as our focus is on supporting the human factors involved rather than directly improving the threat hunting process. However, supporting threat hunters should, in turn, enhance the effectiveness of the activity.

In this section, we present how the literature has characterized threat hunting, including tools, skills, and challenges faced by threat hunters. 
Next, since one of our goals with this study is to provide a model of the process of building a mental model during a threat hunting session, we explore the related literature on the use of models in cybersecurity. We introduce the most popular cognitive models in cybersecurity and their applications in cybersecurity and related fields. Finally, we introduce mental models as an important resource for the development of new tools for cybersecurity.

\subsection{Threat Hunting}
\label{sec_sub_rw_TH}

While the activity of threat hunting has been discussed for many years (since the early 2010s according to Mahboubi~\emph{et al.}~\cite{mahboubi_evolving_2024}), recent studies, such as Nour~\emph{et al.}~\cite{nour2023_th_enterprise}, highlight that threat hunting is still an emerging field with gaps in understanding, particularly in its procedural and organizational aspects. The same authors explain that many organizations remain reactive, responding to alerts and incidents rather than proactively seeking out threats, further underscoring the need for a dedicated threat hunting role.

THs are highly skilled cybersecurity professionals who focus on proactively detecting and mitigating threats, often through methods like log analysis, anomaly detection, and malware analysis~\cite{vanos2018, nour2023_th_enterprise, maxam2024, mahboubi_evolving_2024}. While THs play a key role in the defense of systems and networks, other roles in cybersecurity, such as security analysts and incident responders, can also contribute to threat management, often with a focus on different aspects like monitoring, response, and containment. To support this understanding, we can consult reports for threat hunting surveys that cover the profile of a threat hunter, team sizes and structures, and other aspects based on industry practice \cite{fuchs_sans_2019, fuchs_survey_2023, fuchs2024}.

Threat hunting is performed by analyzing, testing, and evaluating hypotheses based on knowledge gained from a variety of sources~\cite{nour2023_th_enterprise}, such as Security Information and Event Management (SIEM) systems~\cite{mahboubi_evolving_2024, nour2023_th_enterprise}; directly from system, event, and network logs~\cite{nour2023_th_enterprise}; and threat intelligence such as common attacker techniques, tactics, and procedures (TTPs)~\cite{vanos2018} or Indicators of Compromise (IoC)~\cite{nour2023_th_enterprise, maxam2024}.

During structured hunting, generating high-quality hypotheses is paramount~\cite{vanos2018}, and is often a manual process requiring knowledge and expertise~\cite{nour2023_th_enterprise}. However, this is not always the case; unstructured hunting, for example, does not involve the creation of hypotheses~\cite{maxam2024, vanos2018} (we discuss threat hunting processes and workflows further in Subsection~\ref{rw_workflow_th}).

Threat hunters may also belong to broader security teams or larger organizations~\cite{lee2016, vanos2018}. In general, these cybersecurity teams require skills such as an ability to communicate their findings to superiors and shared situational awareness~\cite{mancuso2020}. Situational awareness is the concept of ``knowing what is going on around you''~\cite{endsley_situation_2000}. Much of the research into situational awareness in cybersecurity has a focus on using AI and ML to automate the tasks of security analysts, rather than focusing on supporting the needs of the humans doing the job~\cite{gutzwiller_gaps_2020, mancuso2020, ofte2023}. Additional skills are required by threat hunting teams in particular, for which Hill~\emph{et al.}~\cite{hill2025}, Maxam and Davis~\cite{maxam2024}, and Badva~\emph{et al.}~\cite{badva2024} provide a detailed account---three recent studies conducting interviews in the scope of threat hunting.

Automation plays a significant role in supporting threat hunting activities, and recent advancements—such as the automated generation of attack hypotheses \cite{kaiser2023, nour2024_automa}—have contributed meaningfully to this domain. However, automation alone cannot address all the complexities of threat hunting; human expertise remains essential, as emphasized by different authors such as \cite{nour2023_th_enterprise, mahboubi_evolving_2024, badva2024}. This underscores the importance of a human-in-the-loop approach. Still, despite this importance, many automated tools and related studies often overlook the impact of automation on human interaction and the cognitive dimensions of threat investigation—a gap also highlighted by \cite{badva2024}. 

THs face numerous obstacles that hinder the effective detection and prevention of malicious activities. 
In a recent systematic literature review, Mahboubi~\emph{et al.}~\cite{mahboubi_evolving_2024} summarize key difficulties in threat hunting, including: (a) a lack of labeled data, (b) imbalanced datasets, (c) multiple sources of log data, (d) adversarial techniques, and (e) a shortage of human experts and data intelligence.
These challenges highlight the complexity and ever-evolving nature of cyber threats, as well as the gaps in current methodologies, technologies, and analyst skills, as also noted by other authors \cite{badva2024, maxam2024, hill2025, nour2023_th_enterprise}.
Understanding these challenges is essential to improving the effectiveness and efficiency of threat hunting practices.
In particular, there is a significant opportunity for further research and tool development focused on the behaviors and cognitive processes of threat hunters, which remain underexplored.

\subsection{Models and their Applications in Security}
\label{sec_sub_rw_CMandApp}

In this subsection, we overview various process models used in cybersecurity. For simplicity, we consider everything we discuss to be a \textit{process model,} where we use Curtis \emph{et al.}~\cite{curtis1992}'s definition as (p. 76) ``an abstract description of an actual or proposed process that represents selected process elements that are considered important to the purpose of the model and can be enacted by a human or machine''.

\subsubsection{Models of Threat Actor Behaviors}
Models are used in cybersecurity to make frequently-used attack patterns and vectors more accessible. Three of the most popular threat actor behavior models are the MITRE ATT\&CK~\cite{mitre,mitre2}, Pyramid of Pain~\cite{bianco2013}, and Cyber Kill Chain~\cite{hutchins2011}, which all describe common techniques and behaviors of cyber attackers. The process of a threat hunting team will vary from team to team depending on the different hunting approaches: structured hunting (hypothesis-based) vs. unstructured hunting (data-driven)~\cite{maxam2024, vanos2018}, and the structure of the organization: government vs. private sector~\cite{maxam2024}. 

\subsubsection{Workflow Models for Threat Hunting}
\label{rw_workflow_th}
One of the earliest references to structured processes or workflow models in threat hunting are the works of Gunter and Seitz~\cite{gunter2021} and van Os \emph{et al.}\cite{vanos2018}. Building on foundational concepts such as the Cyber Kill Chain\cite{hutchins2011}, Gunter and Seitz~\cite{gunter2021} propose a six-step cyclical model including \textit{purpose}, \textit{scope}, \textit{equip}, \textit{plan review}, \textit{execute}, and \textit{feedback}. Their model~\cite{gunter2021} emphasizes continuous improvement by ensuring that insights from previous hunts inform future ones. While it aims to provide threat hunters with a standardized methodology and a stronger focus on objectives, the model remains relatively abstract.

In contrast, van Os \emph{et al.}~\cite{vanos2018} offer a more detailed and operationally rich approach through the TaHiTI (Targeted Hunting integrating Threat Intelligence) workflow model. TaHiTI follows a hypothesis-driven, risk-focused methodology and defines three distinct phases. First, in the \textit{Initiate} phase, a trigger (e.g., a security incident, domain knowledge, or input from the MITRE ATT\&CK framework) leads to the creation of a hunting investigation abstract, including an initial hypothesis and priority. Next, the \textit{Hunt} phase involves two key activities: \textit{define/refine}, where the hypothesis is clarified, and \textit{execute}, where data is gathered and analyzed. The \textit{Finalize} phase involves processing results, documenting findings, and handing them off to relevant processes (e.g., incident response, threat intelligence generation, or vulnerability management).

Empirical studies offering interview-based insights into the threat hunting process have only recently emerged (e.g., \cite{maxam2024, badva2024}), introducing a descriptive process instead of a prescriptive approach as in the first two mentioned earlier. 
Maxam and Davis\cite{maxam2024} outline the process across seven distinct stages: \textit{begin hunt}, \textit{mission planning}, \textit{collect intelligence}, \textit{pre-mission activities}, \textit{manual and automatic analysis loops}, and \textit{end mission}, providing detailed contextual information for each phase. 
Conversely, Badva~\emph{et al.}~\cite{badva2024} present a higher-level conceptual model structured as a cyclical workflow. Their workflow model \cite{badva2024} begins with the selection of a \textit{Threat Hunting Method}, which includes approaches as \textit{use-case-based}, predefined scenarios or patterns of suspicious activities to identify and investigate known threats and attack patterns, \textit{intel-based}, leveraging technical threat intelligence, or \textit{random-based} hunting, without a predefined plan. This phase is followed by phases of \textit{Pre-Hunt Planning}, \textit{Data Collection \& Preparation}, \textit{Hunting \& Validation}, and \textit{Remediation \& Reporting}, before looping back to the planning phase.

Together, these models (\cite{gunter2021,vanos2018,maxam2024,badva2024}) offer diverse perspectives on the workflows followed by threat hunters. However, they do not address the underlying cognitive processes---specifically, how threat hunters build, refine, and share their mental models throughout the hunting process.

\subsubsection{Models for Cognitive Tasks}
Other models in cybersecurity focus on cognitive aspects, representing how analysts' minds process information while perceiving, comprehending, and responding to threats. In fields such as cognitive security~\cite{andrade2022} and cyber situation awareness~\cite{barford2010cyber}, these cognitive task models form the basis of proposed strategies to automate the threat hunting process undertaken by cybersecurity analysts~\cite{andrade_cognitive_2019, hurlburt2022, jiang2021, tam2023}. D'Amico \emph{et al.}~\cite{damico2005} previously used a cognitive task analysis approach to construct a generalized workflow of how information assurance analysts build situational awareness and respond to threats. 

Another cognitive task model in this field is the work proposed by Andrade and Yoo~\cite{andrade_cognitive_2019}, providing a comprehensive framework that integrates cognitive science into cybersecurity practices. Their model highlights the integration of technological solutions with the cognitive processes of security analysts, emphasizing the automation of cognitive tasks to enhance efficiency while keeping the human analyst central for critical decision-making. While this model is comprehensive and rich, it is complex and overlooks how important some processes such as building a mental model can be. Our work aims to elucidate critical cognitive processes in threat hunting in a way that has a clear focus and minimizes the visual overload on our own cognitive systems. 

\subsubsection{Models from Other Fields}
Cybersecurity has employed techniques based on popular models originating from other fields, such as the OODA (Observe, Orient, Decide, Act) originally developed by John Boyd~\cite{boyd1996essence} to enhance situational awareness in military aviation operations. The OODA loop is an iterative and adaptive process, crucial for maintaining a competitive edge in dynamic environments such as the cyber threat landscape. OODA's concepts apply to defending from cyberthreats~\cite{muniz2015}, and therefore can also be used in the threat hunting process.

Models from other fields such as in programming~\cite{HEINONEN2023107300, latoza_maintaining_2006, tripathy2015a}, Human-Computer Interaction (HCI)~\cite{Hu2023}, medical education~\cite{Qiao2014}, science education~\cite{rapp_mental_2005}, and business~\cite{Sund2024} can also support the development of better understanding, performance, and decision-making in threat hunting. For example, Heinonen \emph{et al.}\cite{HEINONEN2023107300} and LaToza \emph{et al.}\cite{latoza_maintaining_2006} explore how developers maintain mental models of code, emphasizing communication and implicit knowledge similar to how THs need to share information with their teammates. In medical education, cognitive load theory is used to reduce cognitive load~\cite{Qiao2014}, a challenge also faced by THs. Science education employs mental models to create more intuitive and effective tools~\cite{rapp_mental_2005}. In business, models are used and shared within organizations to improve shared cognition and overall performance~\cite{Sund2024}. These related studies highlight the importance of using models for improving understanding and problem-solving and inspiring the creation of models that support threat hunting.

In the next subsection, we look at related work on mental models in more detail as we explore the construction of TH mental models in our work.

\subsection{Mental Models}
\label{sec_sub_rw_CM}

Mental models are defined by HCI researcher Don Norman as the conceptual models residing in people's minds; they are (typically highly simplified) explanations for how things in the world work~\cite{norman2013}. Mental models form through interactions with the environment, others, and technology~\cite{norman2014}. Mental models are also characterized as evolving over time, often containing inaccuracies, and are limited by the user's technical background or previous experiences~\cite{norman2014}. Two people will not necessarily share the same mental model, and one person may hold multiple mental models of a single item to represent its different functions~\cite{norman2013}.

In the cybersecurity context, existing studies of mental models address how end users perceive threats~\cite{murimi2023, thompson2017}, help cyber analysts conceptualize threat actors and threats themselves~\cite{damico2005}, and enable automation of TH responsibilities.
Mahboubi \emph{et al.}\cite{mahboubi_evolving_2024} highlight how adaptive cognitive processes, including iterative refinement and hypothesis-driven approaches, help analysts address evolving threats, underscoring the importance of mental models in guiding decision-making and integrating human insights with AI tools.

Fortuna \emph{et al.}\cite{fortuna_2023} and Norman~\cite{norman_2017} similarly emphasize the role of cognitive models and human-technology interaction in supporting professionals in complex environments, with a human-centered approach essential for reducing cognitive load and enhancing situational awareness, as also noted by Gutzwiller \emph{et al.}\cite{gutzwiller_gaps_2020}. Murimi \emph{et al.}\cite{murimi2023} stress that effective cybersecurity mental models require a human-centered foundation that accounts for technology, situational awareness, and human behavior. This aligns with insights from HCI, where, as Hu \emph{et al.}~\cite{Hu2023} note, mental models help design more intuitive and accessible systems. Understanding how these models are constructed is crucial for developing tools that truly support their users.

Despite the breadth of related literature, including studies on adaptive cognitive processes~\cite{mahboubi_evolving_2024}, human-technology interaction~\cite{fortuna_2023, norman_2017}, and the need for human-centered foundations in cybersecurity mental models~\cite{murimi2023, Hu2023}, existing research still fails to explain how THs specifically build, use, and share their mental models during active hunting sessions. This gap underscores the need for deeper investigation into the behavioral, cognitive, and collaborative aspects of threat hunting. Addressing this gap, our study focuses on the key cognitive tasks and sequences involved in constructing mental models to navigate complex and dynamic cybersecurity environments.

\section{Study Design}
\label{sec_study_design}

\begin{figure*}[!h]
    \centering
    \includegraphics[width=0.9\linewidth]{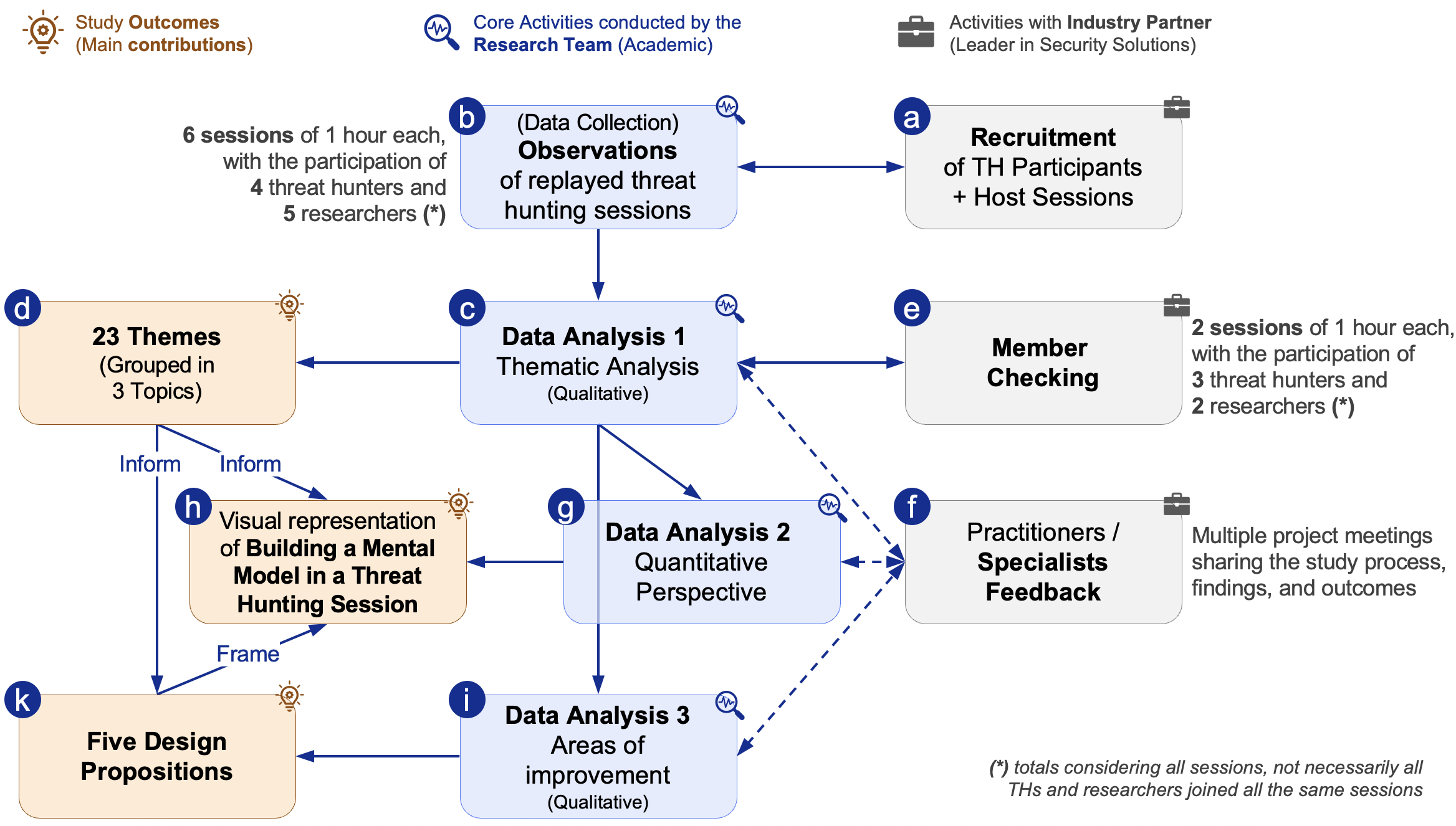}
    \caption{Overview of the study process, main activities conducted by the research team, activities with the industry partner engagement (grey boxes at right side) and the main contributions of the study (orange boxes at the left side). 
    }
    \label{fig:overview_study}
\end{figure*}

In this section, we present the design of our study, structured to investigate the cognitive processes and workflows of THs in a cybersecurity context. Our study design encompasses a series of observation sessions and detailed data analyses to build a comprehensive understanding of the strategies employed by these professionals (which would be challenging to achieve by conducting other methods such as interviews or surveys).

In Subsection~\ref{sec_sub_study}, we provide a concise overview of the main activities of our study. We also introduce the industry partner, the participants, and the protocol we followed to conduct the six observation sessions. 
Next, Subsection~\ref{sec_sub_study_data} explains the approach we employed for our data analysis process.
To conclude, Subsection~\ref{sec_sub_study_validation} covers the member checking and validation strategies used.

\subsection{Study Process and Overview}
\label{sec_sub_study}

We show an overview of our study process in Figure~\ref{fig:overview_study}. In turn, we summarize the main activities performed by the research team, by the industry partner (or with their engagement), and the relation of these activities with the contributions of our study.

A total of six observation sessions were carried out with the participation of four THs and five researchers (see Fig.~\ref{fig:overview_study}-b). While not all THs and researchers attended all the sessions (due to timing constraints), at least two THs and two researchers were present in each session.
It is important to clarify that what we refer to as observations, in the context of our study, means that the research team gathered data through observation of THs who replayed threat hunting sessions (i.e., not real-time hunting as part of their regular work shift).
In the following subsections, we present further details about the study process.

The study protocol was reviewed and approved by the Human Research Ethics Office at the University of Victoria, Canada (project application number 21-0601).

\subsubsection{Industry Partner} 

This study was conducted as part of a research project in collaboration with OpenText.\footnote{https://www.opentext.com/products/cyber-security}
Our industry partner is a global market leader in information management software offering a variety of cybersecurity products and services.
One of the products offered by OpenText is ArcSight Intelligence, described as an advanced threat-detection tool that uses user entity behaviour analytics (UEBA) and unsupervised machine learning models to detect behavioural anomalies across the organization.\footnote{\href{https://www.opentext.com/products/arcsight-intelligence}{https://www.opentext.com/products/arcsight-intelligence}}

Although the study was led by an independent research team, team members from the industry partner were actively engaged during other activities, such as the recruitment and validation of the findings (as explained below).

\subsubsection{Participants} 
We followed an opportunistic recruitment process~\cite{honigmann2003sampling}: the TH participants were collaborators associated with the industry partner of our research project. Our industry partner recruited and selected these four THs based on the research team's request to engage senior THs willing to openly share their work routine during the study (see Fig.~\ref{fig:overview_study}-a). 
The research team did not contact the potential participants directly, and no demographic criteria was used for the selection process (i.e., other than their experience as a threat hunter).
 
The four selected participants were working remotely from different time zones.
To maintain privacy, the research team refrained from capturing any demographic or personal information beyond the first name and email of the participants (which was used for communication on booking the meeting sessions and not associated with their responses). This approach ensured the participants felt secure and respected throughout the study considering the criticality of sharing details about their sensitive threat hunting activities. 

This recruitment style was used due to the challenging nature of finding skilled security professionals willing to share in-depth information considered sensitive in their day-to-day work and that have limited time to offer their expertise due to high demand for their skills as a result of the shortage of trained professionals \cite{sundaramurthy2014}.

\subsubsection{Observation procedure and tools} 
The observation sessions encompass all six formal meetings the research team had with the four participants (see Fig.~\ref{fig:overview_study}-b). 
The participants were encouraged to conduct their threat hunting activities while maintaining their natural environment as much as possible. 
However, in most sessions, the participants chose to share their process based on previous investigations they performed rather than the task planned for their current shift.  
This retrospective approach---replaying a past threat hunting session rather than engaging in real-time hunting---allowed them to present a wider range of cases. 
These cases included hunting sessions triggered by different reasons and hunting approaches (such as use-case or intel-based, see Subsection~\ref{rw_workflow_th} for more details), and not only successful threat discoveries but also their broader investigative process, including marking events that ultimately did not lead to real threats or attacks. This approach reflects how they typically train new team members, making it a representative illustration of their standard practices (though potentially time-constrained).

The observation sessions were conducted online (using Microsoft Teams) and lasted one hour each. The participants kept their video cameras turned off, and the participant leading the session shared their main work screen with the entire group attending. 
No audio or video was recorded during the sessions (as agreed with the participants due to business-sensitive data from clients being shared).

During the sessions, the participants were invited to think out loud while using their regular work tools, which typically include some combination of SIEM and UEBA tooling. For most of the time (session), the participants used a UEBA commercial solution for threat hunting developed by the industry partner (OpenText ArcSight Intelligence).

We attended the observation sessions without predefined hypotheses, allowing us to take notes freely without following a specific template or set of guiding questions (i.e., an exploratory mindset of gathering an understanding of the threat hunting practices and their related cognitive processes).
Additionally, we decided to limit our questions or any interruptions for clarification to only a few to avoid disruptions to the activities and natural flow of the work being conducted by the participants (i.e., participants were not answering predefined questions as would be the case in an interview without the observations). Thus, most of the questions were noted to be clarified later (during validation and member checking sessions, explained in Subsection~\ref{sec_sub_study_validation}).

\begin{figure*}[ht]
    \centering
    \includegraphics[width=1\linewidth]{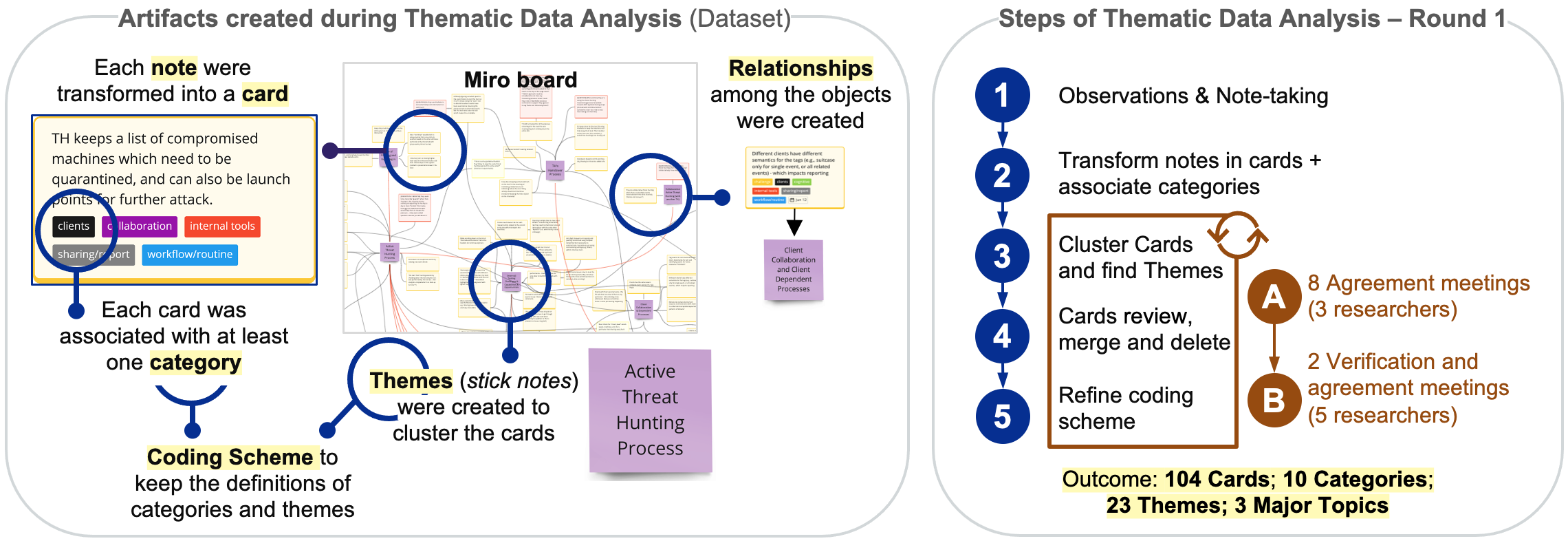}
    \caption{Overview of the thematic data analysis process (Round 1). 
    }
    \label{fig:overview_dataanalysis}
\end{figure*}

\subsection{Data Analysis}
\label{sec_sub_study_data}

Throughout our data analysis process, we documented the steps and decisions made, ensuring transparency in the activities performed. A high-level overview of this process is presented in the following subsections. Focusing on key insights, we present our three rounds of data analysis. The first round entails a thematic analysis \cite{braun_thematic_2022} of the observation sessions (see Fig.~\ref{fig:overview_study}-c). Building on top of the observations notes and results that emerged during Round 1, two more rounds of data analysis were conducted: Round 2, quantitative perspective (see Fig.~\ref{fig:overview_study}-g), and Round 3, identification of areas with problems and opportunities for improvement (see Fig.~\ref{fig:overview_study}-i).

\subsubsection{Round 1: Thematic data analysis}

This round was the most time-consuming and critical part of the data analysis process. 
Figure~\ref{fig:overview_dataanalysis} provides an overview of the artifacts created during the thematic analysis (box on the left) and the main steps conducted in this round (box on the right).

First, each researcher translated their notes from the observation sessions into ``cards'' within an online collaborative board, \textit{Miro}.\footnote{\label{note2}Miro: \href{https://miro.com/}{https://miro.com/}} The cards were then associated with ``categories'' to roughly sort the cards into groups. These activities were done by each researcher individually (corresponding to Steps 1 and 2 in Fig.~\ref{fig:overview_dataanalysis}).

Later, as a group, we evaluated all the cards added to the Miro board. With further analysis, we merged cards with similar items, deleted duplicates, annotated the cards with further details, and adjusted the categories. We then clustered the cards on the Miro board according to the content similarities with other cards, and connected the related cards with relationship arrows. For the emergent clusters, we identified ``themes'' to represent the major concepts within the clustered cards. While analyzing the relationships between the cards and emerging themes, we also kept refining the coding scheme. 
We undertook these activities (corresponding to Steps 3, 4 and 5 in Fig.~\ref{fig:overview_dataanalysis}) in a repetitive manner until the card relationships were stable, and we were in agreement on the themes and categories.

These agreement meetings involved three researchers and were conducted eight times (see Fig.~\ref{fig:overview_dataanalysis}-A). After that, two verification and agreement meetings involving five researchers were conducted to review all cards, categories and themes (see Fig.~\ref{fig:overview_dataanalysis}-B).
Having a smaller team focused on the initial analysis in the first set of agreement meetings made it easier to be consistent and converge on a shared mental model encompassing all the data.  
Having a broader research team involved with the final verification and agreement meetings helped ensure all relevant themes were captured and considered.

It is worth noting that while we employed an inductive approach to analyze the data collected (from the observation notes to the themes), the initial set of categories associated to the cards was derived from the core information concepts identified in our previous study within the threat hunting context (\cite{hill2025,hill2023}).
In the Appendix, we present further details of our coding scheme, including the list of categories (Table~\ref{tab_categories_description}), examples of associated observation notes (cards), and the list of themes (Table~\ref{tab_themes_description}).

At the end of this round, we had 104 cards associated with ten categories and clustered into 23 themes (see Fig.~\ref{fig:overview_study}-d). 
To effectively communicate the findings of this thematic analysis, we further consolidated the 23 themes into three major topics: \textit{supporting construction and communication of the mental model}, \textit{working together}, and the \textit{ability to be effective at finding threats} (presented in Section~\ref{sec_results}).

\subsubsection{Round 2: Quantitative perspective}
In this second round, we continued with an exploratory quantitative analysis approach to delve deeper into the data gathered and the results from the thematic data analysis (Round 1).
For that, we consolidated our study dataset including the cards (i.e., notes from the observation sessions), the coding scheme (i.e., categories and themes), and the metadata created (e.g., relationships among themes)---all the artifacts mentioned in Figure~\ref{fig:overview_dataanalysis}, left side. 

We organized initial guiding questions to expand the understanding of the current themes and explore the cards to find new themes. A few questions were more related to the frequency of occurrence (e.g., which were the themes with more cards associated with?), and others to try to identify any similarity among the artifacts or if any relevant clustering appeared (e.g., how the themes were connected through cards?).
The results of this data analysis round are presented in Subsection~\ref{sec_sub_results_quantitative_analysis}.

\subsubsection{Round 3: Analyzing areas of improvement and creating design propositions}
In the third round of data analysis, our objective was to compile a set of recommendations for designing new tools that could provide cognitive support to THs.

During the observation sessions, we were not concerned with noting ideas or potential solutions to be designed. However, whenever a relevant idea emerged, it was recorded on a card and categorized under ``improvement.'' At the end of Round 1, we had 21 cards categorized as ``improvement'' based on direct observations (such as a desired feature mentioned by a TH) or insights from researchers during agreement sessions regarding problems, opportunities, or improvement ideas.

Using these improvement cards as the starting point for this third data analysis round, we sorted them according to the three major topics (also identified in Round 1, as discussed earlier), which helped reveal relationships between the cards. Subsequently, we colour-coded the cards based on similar areas of improvement, assigning a different colour to each area, such as \textit{Creatively Expanded Search} or \textit{Bookmarks and Note-taking}. This process led to identifying five distinct areas of improvement. These areas of improvement, in addition to our prior design experience, motivated our proposed  five design propositions (see Fig.~\ref{fig:overview_study}-k). These initial design propositions are presented later in Section~\ref{sec_design_propositions}.

\subsection{Validation and Member Checking} 
\label{sec_sub_study_validation}

After the data analysis and the compilation of the main findings into themes (see Fig.~\ref{fig:overview_study}-c), two researchers conducted two member checking sessions with three THs (see Fig.~\ref{fig:overview_study}-e). These sessions were used to clarify questions that emerged from the observation sessions (see Fig.~\ref{fig:overview_study}-b). 

The member checking included only two participants from the observational sessions, as the other two had since left the organization. Additionally, a TH manager (that was not a participant in the observation sessions) was engaged by our industry partner to provide an additional perspective to our member checking as we could only contact two of the observed THs.

Although the member checking process did not impact the mapped themes (see Fig.~\ref{fig:overview_study}-d), new topics became evident during these sessions. We documented these latest considerations and other pending questions requiring new dedicated sessions to be fully understood as future work (i.e., they were assigned as out of scope for this study).

Finally, as part of the validation strategy throughout our study, we shared the preliminary results of each data analysis round with the industry partner during project meetings (see Fig.~\ref{fig:overview_study}-f). These discussions with specialists in the area were fundamental to validating our findings.

\section{Results of the Thematic Data Analysis}
\label{sec_results}

In this section, we present the results of our thematic data analysis, highlighting the comprehensive analysis of threat hunting activities captured during the observation sessions (see Figure~\ref{fig:overview_dataanalysis}). 
To better present the 23 themes emerged during this process, we grouped them into three topics (as major themes): \textit{support construction and communication of mental model} (Subsection~\ref{sec_sub_results_theme1}), 
\textit{working together} (Subsection~\ref{sec_sub_results_theme2}), and 
\textit{the ability to be effective at finding threats} (Subsection~\ref{sec_sub_results_theme3}). 
Table~\ref{tab:themes} shows these three topics and the respective themes associated with each of them.

We present the themes without any differentiation by relevance or frequency of occurrence, but we reflect on this during a second round of data analysis, which is summarized in Subsection~\ref{sec_sub_results_quantitative_analysis}.

As part of the theme description, we add examples of the observation notes that support the theme creation. Thus, it is worth reminding that some of the themes are closely related to SIEM and UEBA tools context and THs working in teams---subject of discussion later in the paper (Section~\ref{sec_discussion}).

\begin{table}[htbp]
    \centering
    \caption{Thematic data analysis Results: 23 themes grouped into three major topics}
    \label{tab:themes}
    \begin{tabular}{p{0.4cm}p{7cm}}
        \toprule
        \multicolumn{2}{c}{\shortstack{Topic 1: Support Construction and Communication of Mental Model}} \\ 
        \midrule
        \ref{sec_sub_themes_1_1} & Active Threat Hunting Process \\
        \ref{sec_sub_themes_1_2} & Frequent Use of Memory \\
        \ref{sec_sub_themes_1_3} & Internal to External Data Linking \\
        \ref{sec_sub_themes_1_4} & Technical Skills and Experience \\
        \ref{sec_sub_themes_1_5} & Significant Event Marking  \\
        \ref{sec_sub_themes_1_6} & Feedback Loop between TH and Tool During an Active Threat Hunt \\
        \ref{sec_sub_themes_1_7} & Ease of Pivoting and Exploring in UIs \\
        \ref{sec_sub_themes_1_8} & Attacker Strategy \\
        \ref{sec_sub_themes_1_9} & Mental Model of Active Threat Hunt Activity \\ 
        \ref{sec_sub_themes_1_10} & Mental Model of Client’s System\\              
        \midrule
        \multicolumn{2}{c}{\shortstack{Topic 2: Working Together}} \\ 
        \midrule
        \ref{sec_sub_themes_2_1} & Reporting \\
        \ref{sec_sub_themes_2_2} & Client Collaboration and Dependent Processes \\
        \ref{sec_sub_themes_2_3} & Collaboration with Threat Hunting Tool Maintainers and Developers \\
        \ref{sec_sub_themes_2_4} & Collaborative Active Threat Hunting \\
        \ref{sec_sub_themes_2_5} & Handover Process \\
        \ref{sec_sub_themes_2_6} & Documentation of Active Threat Hunting Findings \\
        \midrule
        \multicolumn{2}{c}{\shortstack{Topic 3: Ability to be Effective at Finding Threats}} \\         
        \midrule
        \ref{sec_sub_themes_3_1} & Information Resource Challenges \\
        \ref{sec_sub_themes_3_2} & Internal Tooling Capabilities, Challenges, and Opportunities \\
        \ref{sec_sub_themes_3_3} & Event Search Capabilities \\
        \ref{sec_sub_themes_3_4} & When to Stop Hunting? \\
        \ref{sec_sub_themes_3_5} & Limitations of UEBA \\
        \ref{sec_sub_themes_3_6} & Missing Pivot Points Between Correlated Events or Groups of Events \\
        \ref{sec_sub_themes_3_7} & Data Availability Limitations \\
        \bottomrule
    \end{tabular}
\end{table}

\subsection{Supporting Construction and Communication of Mental Model}
\label{sec_sub_results_theme1}
This first topic, \textit{supporting construction and communication of mental model}, encompasses ten themes (described in the following subsections). It involves aspects of the cognitive mental model (of the TH) as an essential artifact from the hunt and so its importance to be supported.

\subsubsection{Active Threat Hunting Process} 
\label{sec_sub_themes_1_1}
This theme refers to the workflow or routine related to the active threat hunting process (e.g., steps and checklists). It is also associated with the structures, standards, and objectives for a particular TH team, such as using heuristics, guidelines, or checklists.
Observation note: \textit{Example threat hunting workflow: first, identify the entry point of the attack (``how did the authentication start?''), next, ``identify the activities that the attacker did with their access''.}

\subsubsection{Frequent Use of Memory} 
\label{sec_sub_themes_1_2}
This theme refers to using the TH's memory to store helpful information, necessary or relevant for active threat hunting purposes, such as process names and status codes.
Observation note: \textit{A TH remarked, ``I think I made a mistake,'' while copying and pasting some numerical values/tags from OneNote into \THtool.}

\subsubsection{Internal to External Data Linking}
\label{sec_sub_themes_1_3}
This theme refers to linking internal threat hunting data to external resources, for example, linking Windows process names in event logs to their place in the online documentation.
Observation note: \textit{THs use multiple internet search engines to check different codes and info. For example, a TH first used Bing before resorting to Google when the result was unsatisfactory.}  

\subsubsection{Technical Skills and Experience}
\label{sec_sub_themes_1_4}
This theme refers to THs' technical skills and knowledge background, including operating systems, system administration, computer networks, and other areas.
Observation note: \textit{THs frequently rely on memory. For example, a TH remarked `If I remember correctly' while investigating the use of a specific executable.}

\subsubsection{Significant Event Marking}
\label{sec_sub_themes_1_5}
This theme refers to how and why events are annotated. For instance, the ``how'' can be the tool tags (i.e., a visual icon) and the ``why'' the bookmarks and communication with the client.
Observation note: \textit{THs add tags and comments within \THtool~for the next TH only in the case of a potentially critical anomaly. They do not keep notes for events they investigate but conclude are not anomalous, such as their reasoning for dismissing an event.}

\subsubsection{Feedback Loop Between a TH and Their Tool During an Active Threat Hunt}
\label{sec_sub_themes_1_6}
This theme refers to the TH's ability to inform the tool of their findings and next steps. It is also related to the tool's ability to support the TH using this additional context (e.g., by filtering, suggesting or highlighting information and views).
Observation note: \textit{How to make the ML more supportive (proactive) to TH activities?} 

\subsubsection{Ease of Pivoting and Exploring in UIs}
\label{sec_sub_themes_1_7}
This theme refers to UI-specific challenges, ideas, and improvements related to how the TH pivots on events. It is connected to ``Missing Pivot Points Between Correlated Events or Groups of Events'' theme, which describes data processing rather than presentation. This theme is purely related to presentation and user interaction.
Observation note: \textit{The TH keeps the view of risky machines open on the left to pivot on in case a suspicious entity is discovered.}

\subsubsection{Attacker Strategy}
\label{sec_sub_themes_1_8}
This theme describes the TH's processes for tracing the activities and movement of an attacker through the system and learning the patterns of the attacker to find related events or activities.
Observation note: \textit{An important consideration is the TH's ability, once an attacker's strategies are known, is to be able to use something like \THtool's ``find similar'' feature for clustering for future searches.}

\subsubsection{Mental Model of Active Threat Hunt Activity}
\label{sec_sub_themes_1_9}
This theme refers to the internal (in the individual’s head) or external (in software, on paper, etc.) organization or conceptual model the TH builds of notable or suspicious events. It also includes the story line (or timeline) of these events. This is a creative and subjective process compared to the ``Active Threat Hunting Process'' theme. 
Example of an observation note: \textit{The TH must create a 'big picture' of the attack, for themselves, other THs, management, and others.} 

\subsubsection{Mental Model of Client’s System}
\label{sec_sub_themes_1_10}
This theme refers to the internal (in the individual's head) or external (in software, on paper, etc.) organization or conceptual model the TH builds of the client's environment. This mental model contextualizes the TH's activities and understanding and provides the TH with their bearings and intuition during the hunt.
Observation note: \textit{How do THs build an understanding of the client's system? (Long-term mental model).}

\subsection{Working Together}
\label{sec_sub_results_theme2}
This second topic, \textit{working together}, groups six themes (described in the following subsections). It involves aspects of how THs collaborate with others, document, report, and share their findings.

\subsubsection{Reporting} 
\label{sec_sub_themes_2_1}
This theme relates to techniques, tools, and processes used (by anyone, e.g., THs or clients) to generate and communicate reports on threat hunting activities, such as findings and results.
Observation note: \textit{\THtool~generates custom reports based on the flags/tags created for events.}

\subsubsection{Client Collaboration and Dependent Processes}
\label{sec_sub_themes_2_2}
This theme relates to TH interactions with their clients and processes that are unique or dependent on the clients for whom the process is being conducted.
Observation note: \textit{Behavioral analysis by the TH during a hunt requires pre-existing communication with client to understand acceptable/expected patterns of behavior.}

\subsubsection{Collaboration with Threat Hunting Tool Maintainers and Developers} 
\label{sec_sub_themes_2_3}
This theme relates to processes through which a TH can effect change in their tools through the tool's maintainers and developers.
Observation note: \textit{THs work with the data science team to update the activity patterns that \THtool~can detect and consider, improving tool's ability to accurately classify risk.}

\subsubsection{Collaborative Active Threat Hunting} 
\label{sec_sub_themes_2_4}
This theme relates to active threat hunting, performed with real-time communication and collaboration with other THs looking at the same data. By active, we mean, what is characterized by TH action rather than by contemplation or speculation.
Observation note: \textit{THs hunt together when there are multiple events associated with the same anomaly (``divide and conquer'').}

\subsubsection{Handover Process}
\label{sec_sub_themes_2_5}
This theme relates to the protocol and resources THs use to hand over information during shifts or during the shift handover process.
Observation note: \textit{THs keep notes for the next TH using OneNote. Information includes the date/time and links to pages in \THtool. The hand-over process varies from one TH to another (sometimes meetings, but mostly just providing the OneNote).}

\subsubsection{Documentation of Active Threat Hunting Findings} 
\label{sec_sub_themes_2_6}
This theme relates to the information recorded during threat hunting activities and how that information is created, represented, used, applied, and shared.
Observation note: \textit{Findings and logs from each hunt are shared with the next TH through OneNote.}

\subsection{Ability to be Effective at Finding Threats}
\label{sec_sub_results_theme3}
This third topic, \textit{ability to be effective at finding threats}, groups seven themes. It involves aspects of how THs use the available tools to support themselves in searching and gathering relevant information (among other tasks) to facilitate their hunting process. From the discussion of this third topic, insights on complementary strategies to support the TH in being effective emerged. 

\subsubsection{Information Resource Challenges} 
\label{sec_sub_themes_3_1}
This theme relates to challenges associated with using information resources such as websites and documentation. It applies to static information sources (static resources being websites and forums and dynamic resources being data logs that are used in the SIEM tools) and is not associated with compute resources.
Observation note: \textit{The threat hunting team has an internal dictionary, but it is not always up to date, meaning they mostly rely on the internet. For example, the CrowdStrike data dictionary is outdated.}

\subsubsection{Internal Tooling Capabilities, Challenges, and Opportunities} 
\label{sec_sub_themes_3_2}
This theme relates to limitations, inefficiencies, ideas, and strong points in the TH's current tools. It only applies to cards specific to the internal tool (design/UI) that are also not strongly related to other themes. This theme acts as a catch-all for cards related to the internal tool without a specific theme. For cards that include tooling challenges but also relate to other themes, the ``internal tools'' category (tag) was applied instead, and an explicit edge to this theme was omitted.
Observation note: \textit{THs relied on using multiple browser tabs to view details for separate events. These tabs were all labeled simply ``Explore'' in the browser, making it seemingly difficult to manage tabs or navigate between event details easily.}

\subsubsection{Event Search Capabilities} 
\label{sec_sub_themes_3_3}
This theme relates to (tooling) capabilities to support THs in searching for events, e.g., using filters or keywords.
Observation note: \textit{More customizable search/filters/sorts would be useful. For example, allowing filtering based on entity anomaly score delta.}

\subsubsection{When to Stop Hunting?} 
\label{sec_sub_themes_3_4}
This theme relates to the heuristics or frameworks used by THs to decide when to conclude the active threat hunt.
Observation note: \textit{THs feel pressured when deciding when to stop hunting. It may help to include features (e.g., timer, percentage coverage, or checklists) that support the TH in deciding when to conclude the hunt.}

\subsubsection{Limitations of UEBA}
\label{sec_sub_themes_3_5}
This theme relates to the limitations inherent to anomaly detection approaches in User and Entity Behavior Analytics (UEBA)---attackers can create noisy activity to reduce the likelihood of any of their malicious activity being flagged as ``anomalous''.
Observation note: \textit{A TH remarked that ``attackers hide attacks by creating lots of noise,'' which makes the anomaly scores less useful and potentially misleading. In such cases, THs become more reliant on raw event search capabilities to find anomalous events.}

\subsubsection{Missing Pivot Points Between Correlated Events or Groups of Events} 
\label{sec_sub_themes_3_6}
This theme relates to the missing ability to pivot on events to detect anomalies across related entities (e.g., navigate to similar events or entities instead of individual entities). It is distinct from the UI issues as it is purely the quality/existence of these associations after processing by the backend.
Observation note: \textit{If a TH finds a suspicious event (in \THtool), they must manually ``zoom-out'' in the data to look for events that were pinned as suspicious in the past.}

\subsubsection{Data Availability Limitations} 
\label{sec_sub_themes_3_7}
This theme relates to problems with missing data, and not in how the data is being processed or analyzed, which is strictly a failure of the tool performing the analysis.
Observation note: \textit{A lack of data from the client or lack of visibility into the client's system is a challenge for THs, as it means ``you don't have a map to navigate.''}

\subsection{Expanding on Thematic Analysis Findings}
\label{sec_sub_results_quantitative_analysis}

After completing our thematic data analysis in round one using a qualitative approach and before compiling our proposed design propositions (detailed in Section~\ref{sec_design_propositions}), we performed a second round of data analysis using a quantitative approach. 
In the description of this approach, we again refer to cards, categories and themes, as we defined in Figure~\ref{fig:overview_dataanalysis}.
In this second round, we formulated initial questions to guide the analysis. For example, \textit{What are the top themes with the highest number of associated cards?} (see Table~\ref{tab_top_themes}) and \textit{What are the top categories associated to the cards?} (see Table~\ref{tab_top_categories}).

\begin{table}[htbp]
    \centering
    \caption{What are the top themes with the highest number of associated cards?}
    \label{tab_top_themes}
    \begin{tabular}{{p{0.7cm}p{5.2cm}p{1cm}}}
        \toprule
        \centering \textbf{Top \#} & \textbf{Theme Name} (Section) & \textbf{\# Cards} \\
        \toprule
        \centering 1 & Internal Tooling Capabilities, Challenges, and Opportunities (\ref{sec_sub_themes_3_2}) &  14 \\
        \midrule        
        \centering 2 & Active Threat Hunting Process (\ref{sec_sub_themes_1_1}) & 12 \\
        \midrule        
        \centering 3 & Mental Model of Active Threat Hunt Activity (\ref{sec_sub_themes_1_9}) & 11 \\
        \midrule        
        \centering 4 & Client Collaboration and Dependent Processes (\ref{sec_sub_themes_2_2}) & 10 \\
        \midrule        
        \centering 5 & Handover Process (\ref{sec_sub_themes_2_5}) & 9 \\
        \midrule        
        \centering 6 & Internal to External Data Linking (\ref{sec_sub_themes_1_3}) & 8 \\
        \midrule        
        \centering 7 & Event Search Capabilities (\ref{sec_sub_themes_3_3}) & 8 \\
        \midrule        
        \centering 8 & When to Stop Hunting? (\ref{sec_sub_themes_3_4}) & 8 \\
        \midrule        
        \centering 9 & Significant Event Marking (\ref{sec_sub_themes_1_5}) & 6 \\
        \midrule        
        \centering 10 & Reporting (\ref{sec_sub_themes_2_1}) & 5 \\
        \bottomrule
    \end{tabular}
\end{table}

\begin{table}[htbp]
    \centering
    \caption{What are the top categories associated to the cards?}
    \label{tab_top_categories}    
    \begin{tabular}{{p{0.7cm}p{2.5cm}p{1cm}}}
        \toprule
        \centering \textbf{Top \#} & \textbf{Category Name} & \textbf{\# Cards} \\
        \toprule
        \centering 1 & Internal\_Tools & 81 \\
        \midrule
        \centering 2 & Workflow/Routine & 80 \\
        \midrule        
        \centering 3 & Cognitive & 55 \\
        \midrule
        \centering 4 & Challenge & 40 \\
        \midrule        
        \centering 5 & Collaboration & 33 \\
        \midrule        
        \centering 6 & Efficiency & 30 \\
        \midrule        
        \centering 7 & Improvement & 21 \\
        \midrule        
        \centering 8 & External\_Tool & 19 \\
        \midrule        
        \centering 9 & Clients & 17 \\
        \midrule        
        \centering 10 & Sharing\_Report & 14 \\
        \bottomrule
    \end{tabular}
\end{table}

Additionally, when analysing the categories that often appear on the same cards, we noticed some interesting relations, for instance, a common relationship between ``workflow/routine''  is ``cognitive'', which suggested to us that the team's workflow is heavily influenced by cognitive factors or vice versa. Another example, for ``cognitive'' category, the most related category was ``efficiency'', which suggested that improvements to cognitive load could lead to improvements in efficiency (matrix of categories similarity is available in Appendix-Figure~\ref{fig:categories_similarity}).

Although this second round of data analysis did not uncover new themes, it was through this process that our research team had some insightful discussions. These thoughtful discussions led to evaluating patterns and trends on the data gathered and the results, from which the \THmodel~emerged (as noted in Figure~\ref{fig:overview_study}-h and introduced next, Section~\ref{sec_sub_results_model}).

\begin{figure*}[h]
    \centering
    \includegraphics[width=1\linewidth]{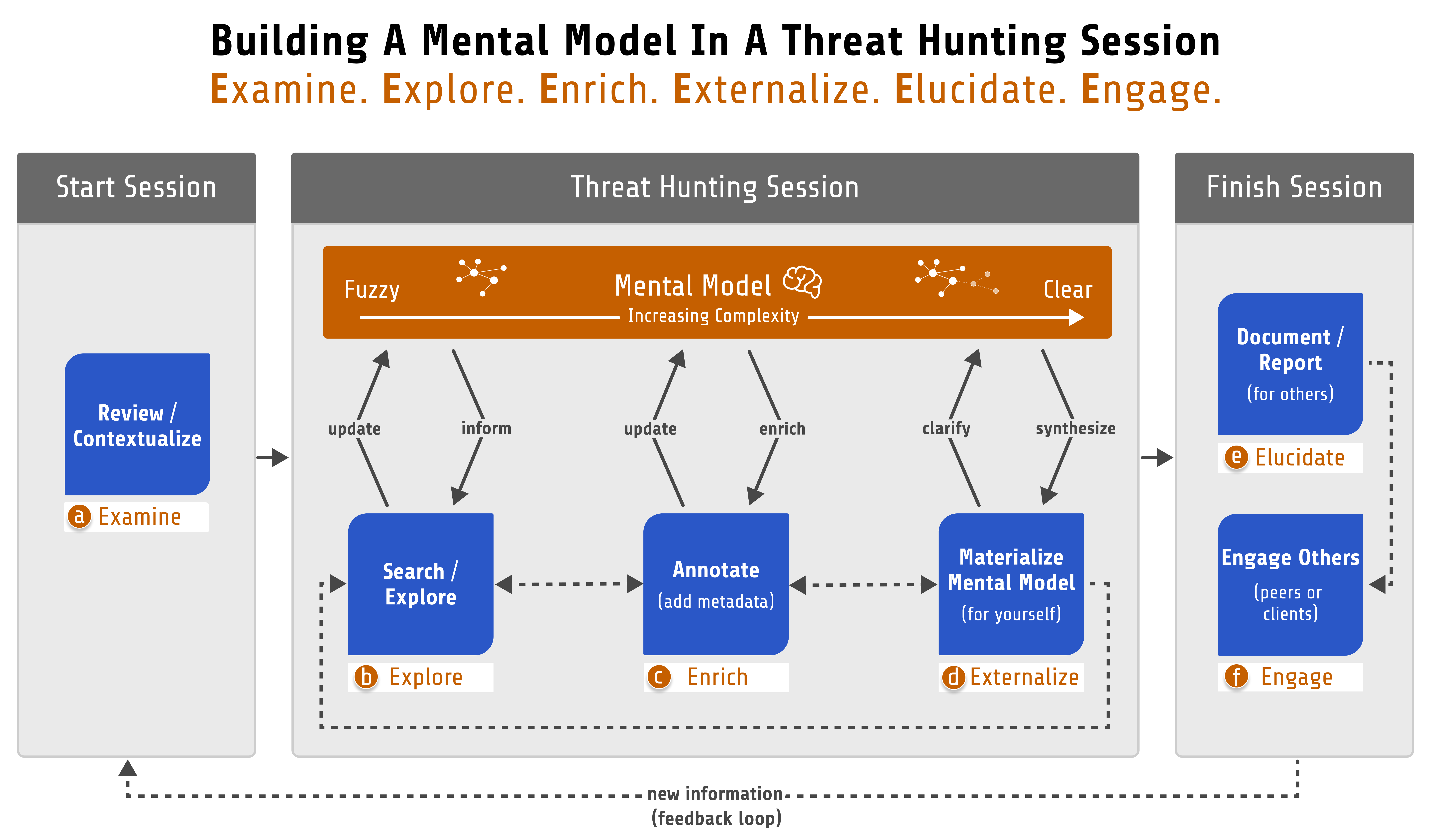}
    \caption{The threat hunter process of building a mental model during a threat hunting session. This visual representation includes a total of six intentions---(a) \textit{Examine}, (b) \textit{Explore}, (c) \textit{Enrich}, (d) \textit{Externalize}, (e) \textit{Elucidate}, and (f) \textit{Engage}---and six related activities (blue boxes positioned above the intentions) organized in three phases (grey boxes labelled with \textit{Session}). We highlight the \textit{Mental Model} to capture the complex, often unseen, cognitive processes central to threat hunting.} 
    \label{fig:model}
\end{figure*}

\section{Building a Mental Model in a Threat Hunting Session}
\label{sec_sub_results_model}

Our findings reveal that THs rely heavily on their ability to construct and refine mental models to navigate complex investigations. In this section, we delve into the critical process of building a mental model in threat hunting sessions, highlighting its significance in THs' cognitive processes.

Figure~\ref{fig:model} shows our proposed \THmodel, which is structured around three phases: (I) initiating the hunting session, (II) the main part of the threat hunting session, and (III) the conclusion of the hunting session. For each phase, activities and intentions that guide the TH work processes, are summarized. Table~\ref{model_activities_intentions} presents the six intentions and activities.

\begin{table}[h]
    \centering
    \caption{List of intentions and activities per phase}
    \label{model_activities_intentions}    
    \begin{tabular}{{p{0.8cm}p{1.7cm}p{3.2cm}}}
        \toprule
        \centering \textbf{Phase} & \textbf{Intention} & \textbf{Activity}\\
        \toprule
        \centering I & a. Examine & Review / Contextualize \\
        \bottomrule        
        \centering II & b. Explore & Search / Explore \\
        \centering  & c. Enrich & Annotate \\     
        \centering  & d. Externalize & Materialize Mental Model \\  
        \bottomrule        
        \centering III & e. Elucidate & Document / Report \\  
        \centering  & f. Engage & Engage Others \\          
        \bottomrule
    \end{tabular}
\end{table}

To provide a walk-through of the \THmodel, we illustrate a threat hunting session performed by a TH named Olivia.\footnote{Olivia is inspired by the persona reported in \cite{hill2025}.} Olivia is an experienced TH working as an external consultant who uses the same SIEM tool as her clients. Olivia is also the team lead of a group of THs who share the same portfolio of clients. 

Olivia starts her threat hunting work shift by examining the notes and the team's checklist document of the last threat hunting session. She reviews the details and contextualizes her investigation scope with the information she receives during the handover process (i.e., with the TH team covering the previous shift). During this activity (see Fig.~\ref{fig:model}-a), Olivia \textit{examines} the problem, question, or situation that lacks clarity before starting her investigation within the threat hunting session.

Based on the input from the first phase, Olivia now embarks on the second phase; the hunting phase where the actual investigation takes place. During the hunting session phase, she builds a \textit{mental model} of possible threats in the client's network. The mental model represents what happens in the individual’s head~\cite{norman2013}, i.e., their cognitive effort or memory).

First (see Fig.~\ref{fig:model}-b), Olivia \textit{explores} the data by searching, filtering, and moving through the data available on her SIEM tool. She compares entities (i.e., user, machines, IP addresses, files, or others) from a particular user against their observed behaviour and also their peers while questioning herself ``is this user supposed to do this?'' or ``is it normal for this user to work on weekends?''. If yes, she asks herself ``is this activity or behaviour within the expected scope?''.

Next (see Fig.~\ref{fig:model}-c), Olivia \textit{enriches} the data by annotating the entities and adding metadata to the telemetry and event data to be used by herself and others (such as other THs in her team or her clients). 
Olivia's SIEM tool allows her to add comments and marks (tags or flags) to indicate data that needs further attention. She follows the standard conventions aligned with her team and clients that have access to the same data. For example, after finding something interesting, she adds a ``pin'' mark to an event for another TH to continue the investigation later. She adds a ``briefcase'' mark to an event that is worth being reported to a client, and adds a ``fire'' mark to an event that refers to a threat or an attack that is ongoing. 

Following (see Fig.~\ref{fig:model}-d), Olivia \textit{externalizes} her mental model by creating notes or saving links to the pieces of evidence she considers relevant to synthesize her hunting process. Sometimes, she creates diagrams, such as a mind map, to visualize the computer network and which machines are compromised. Externalizing this information allows her to reduce her reliance on memory and clarify her thoughts. Compared to the (b) \Enrich~intention, her externalization efforts are not meant to be shared with others---the externalization is for her benefit, so there are no conventions to worry about (i.e., it is whatever representation works best for Olivia).

It is also important to clarify that the three activities performed during the core threat hunting session phase (see Fig.~\ref{fig:model}-b-c-d) can happen multiple times in any order, as new evidence is iteratively uncovered during the hunting process.  

As the outcome of the second phase, Olivia has clarity about what's happening on the client's network, and an improved understanding of the investigation scope. She has transitioned from a mental state of \textit{fuzziness} to \textit{clarity} while building her mental model during the hunting session. This enables her to confidently finish the hunting session, and enter the third (and final) phase.

When Olivia's threat hunting session ends, either because her shift ends or because she has gathered enough information to feel confident in her findings for the particular investigation scope, she conducts two final activities with her newfound knowledge. 
First (see Fig.~\ref{fig:model}-e), Olivia \textit{elucidates} the findings by creating reports and documentation that clarifies the important information from her threat hunting session. She prepares notes for the handover meeting, where the next TH will be briefed on the ongoing investigation. Additionally, she creates a report about suspicious events to be shared with the client, and a comprehensive overview of her threat hunting session.

Finally (see Fig.~\ref{fig:model}-f), Olivia \textit{engages} the client security team and requests their involvement in reviewing the documentation she compiled.  She asks for confirmation regarding the questions she raised about the user behaviour that could lead to a potential threat. Before ending her threat hunting session, she also raises the alarm with her team about a possible ongoing attack. 

This last activity (\textit{Engage Others}) primarily involves the TH initiating contact with others as part of a collaborative effort.  The engagement activity might be with a client, or with her peers as the start of a shift handover process. 
Regardless of the type of engagement, the TH's activities in this final phase can trigger new challenges or questions, which serve as input to reinitiate the investigation process, creating a \textit{feedback loop}. 

We consider our \THmodel~representation to be abstract but detailed enough to help inform strategies and inspire innovation to support the THs. 
We further discuss the implications of our model in Subsection~\ref{sec_sub_discussion_implications}. In the next section, we present five design propositions that emerged from our study.

\section{Design Propositions}
\label{sec_design_propositions}

This section introduces five design propositions (DP) that emerged from our comprehensive data analysis (explained in Section~\ref{sec_study_design}). 
These propositions, informed by our experience as tool designers, aim to support the intentions of the threat hunter's process of building a mental model (see Figure~\ref{fig:overview_study}-k). The five propositions we propose include:
\begin{itemize}[\itemindent=18pt]
    \item [DP 1]\textit{Creating a Story or Timeline of Events}
    \item [DP 2] \textit{Visualizing / Navigating Connections (Spatial)}
    \item [DP 3] \textit{Creatively Expanded Search}
    \item [DP 4] \textit{Waypoints and Note-Taking}
    \item [DP 5] \textit{Integrating External Resources}
\end{itemize}

\begin{figure}[h]
    \centering
    \includegraphics[width=1\linewidth]{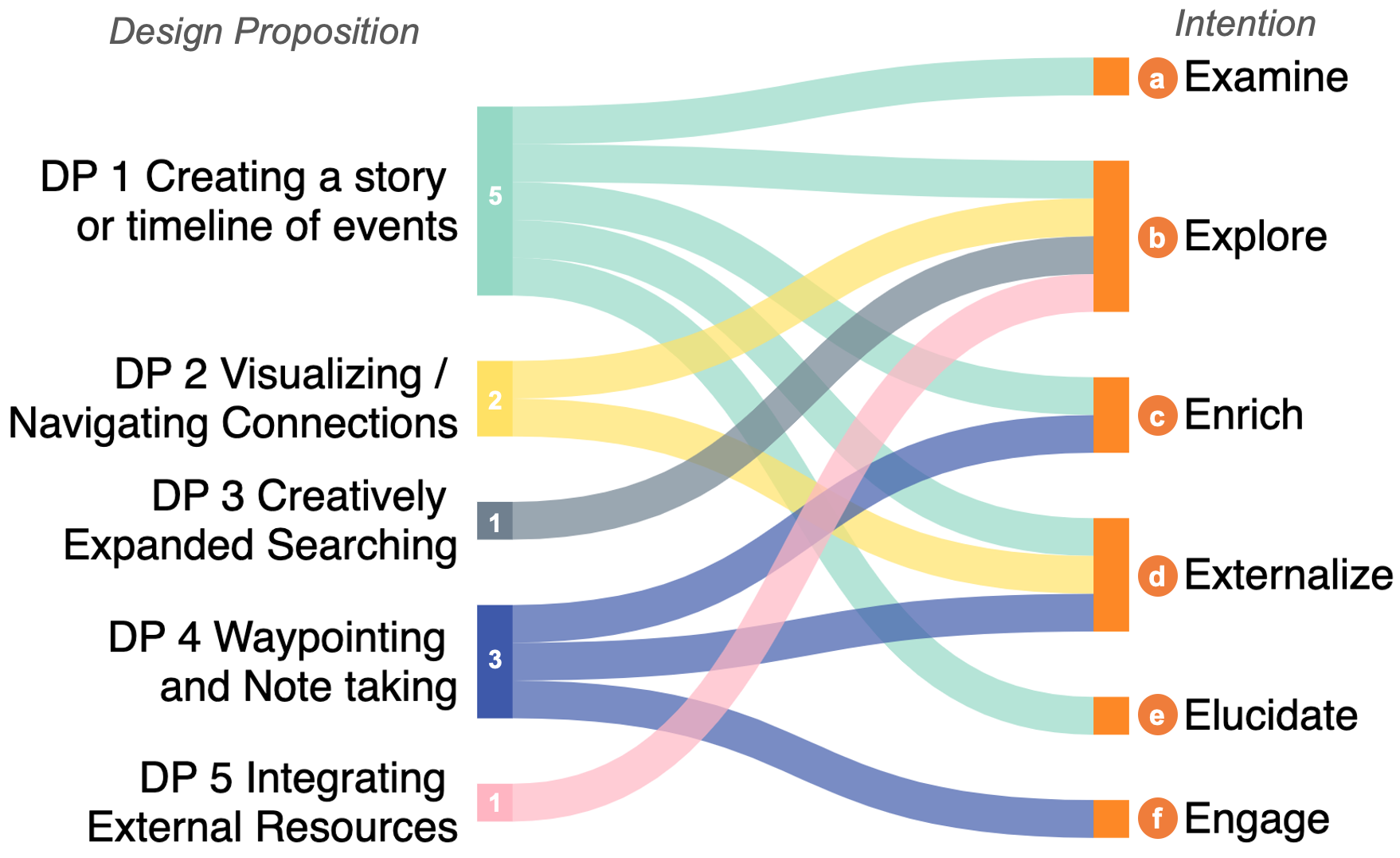}
    \caption{Relation of the five design propositions and the six intentions presented in the \THmodel.}
    \label{fig:ideas_model}
\end{figure}

Figure~\ref{fig:ideas_model} shows the relation between these DP and the different intentions (or their associated activities) of the \THmodel.

In the next subsections, the five DP are presented at a high level but with what we believe to be sufficient detail to convey their potential impact. Although some DPs are more specific or detailed than others, for each DP, we provide the goal, problems, potential solutions (or features), and user stories. 
For the examples of user stories, we refer to the TH personas created by Hill et al.~\cite{hill2025}: Olivia (the creative team lead TH who hunts proactively and prides herself in her leadership skills), Jay (the newer TH, fresh from an academic track with excellent analytic skills and a reactive hunting style), and Thomas (the most experienced TH that works in a small team with an intuitive hunting style).

It is important to note that aspects regarding usability or friction points in the current UI of \THtool~were not the target of our study. Still, some observations were made about the UI friction points in the presentation of the findings in Section~\ref{sec_results}. Our study is also not intended to be an exhaustive or definitive set of DP based on our observations, analysis, and discussions. Therefore, we selected the five most prominent DP to inspire continued exploration. We are confident these DP provide a foundation for innovation and advancement to support the threat hunting cognitive process.

\subsection{Creating a Story or Timeline of Events}

DP 1 \textit{Creating a Story or Timeline of Events} aims to support the TH's process of formulating a clear story of what happened during a possible attack, as the TH's mental model goes from an initially fuzzy state, to a clear understanding, allowing the TH to track progress, update their mental model, and collaborate with their team.

Related problems include:
(a) many disconnected threads of information to keep track of;
(b) high cognitive load while hunting; and
(c) THs relying on their memory to store information.

Potential solutions could consider:
\begin{itemize}
    \item A mind map integrated into \THtool~that allows the TH to keep track of emerging connections and findings.
    \item A timer to know how long a hunt has taken so far.
    \item Percentage coverage indicator to track the progress of what has been reviewed.
    \item A checklist to ensure critical hunting tasks are completed.
    \item The ability to share a partial understanding with a mind map to collaborate with peers on a hunt.
    \item Collect information leads into story frame buckets (abstract container) with notes and annotations that can be organized into a story.
    \item Find information quickly with search filters that are driven by the contents of the story frames.
\end{itemize}

User stories (a more comprehensive list of user story examples for DP 1 is provided in Appendix-Table~\ref{tab_dp_stories}):
\begin{itemize}
    \item As Olivia, I want to externalize (draw) the story of the hunt I am working on so that I can clarify my thoughts and reduce my cognitive load.
    \item As Thomas, I want to add notes to my externalized mental model so that I can explain the data to myself and others.
    \item As Olivia, I want to share my externalized mental model with other THs to align our mental models and share our findings during an active hunt.
\end{itemize}

\subsection{Visualizing and Navigating Connections}

DP 2 \textit{Visualizing / Navigating Connections (Spatial)} goal is to orient the TH to the map of the client's system/network, help THs navigate to related connections while maintaining bearings (orientation), and let THs be able to sense of where they are within the map. 

Related problems include:
(a) difficult to maintain a sense of orientation (e.g., many open tabs in the web browser);
and (b) unable to navigate along some related connection paths without re-searching.

Potential solutions could consider:
\begin{itemize}
    \item A spatial visual map of the client's network as an orientation tool.
    \item A way to pivot from one entity to another based on the entity’s interactions.
    \item A way to see what happened before/after an event.
    \item A way to navigate between event details easily and keep track of where you are.
\end{itemize}

User stories (a more comprehensive list of user story examples for DP 2 is provided in Appendix-Table~\ref{tab_dp_stories}):
\begin{itemize}
    \item As Thomas, I want to visualize and see where I am within a spatial map of the client's system/network so that I can orient more easily to understand the activity I'm seeing.
    \item As Jay, I want to navigate from one entity to other related entities based on interactions in the data, so that I can explore the connections and make sense of what's happening.
\end{itemize}

\subsection{Creatively Expanded Search}

DP 3 \textit{Creatively Expanded Search} goal refers to patterns in how THs seek information and the possibility of constructing an extended search capability that matches the TH's intuitive direction of inquiry.

Related problems include:
(a) TH wants to see if there are similar execution patterns happening across a variety of machines when seeing a suspicious execution pattern, and have no direct way to support this inquiry;
and (b) TH wants a way to quickly filter subsets of data (based on a variety of different things).

Potential solutions could consider:
\begin{itemize}
\item Search pathways that can find events by their similarity to an execution pattern, possibly by fingerprinting or clustering similar patterns up front.
\item The ability to save execution patterns that represent common attack patterns as they are learned.
\item The ability to search across machines by execution/attack pattern.
\end{itemize}

User stories (examples for DP 3):
\begin{itemize}
    \item As Jay, I want to search for patterns that are similar to known attack strategies and be able to ``find similar'' patterns to one I'm seeing.
    \item As Thomas, I want to search for patterns in the data that are similar to observed patterns in the attacker's activities so far, so that I can discover the full scope of compromised machines.
\end{itemize}

\subsection{Waypoints and Note-Taking}

DP 4 \textit{Waypoints and Note-Taking} refers to TH being able to add metadata and notes that support the construction of the story of the threat that is unfolding in their mental model. 

Related problems include:
(a) TH needs the ability to take more notes within the TH tool;
(b) TH loses notes or bookmarks during the threat hunting;
(c) TH misses what other THs did during the same investigation;
and (d) difficulty of sharing notes and bookmarks to inform future hunts.

Potential solutions could consider:
\begin{itemize}
    \item Bookmarking events and entities within the tool to inform future hunts (not only for documentation).
    \item Supporting waypoints that are constructed during navigation that help the TH visualize their ``journey'' through the data, and return to previous waypoints.
    \item Rank or like (or dislike) previous comments to support their agreement with a hypothesis, or say something is no longer critical.
    \item Keeping notes within main TH tool to support future hunts (not just documentation).
    \item Supporting a thread of comments that link through multiple events.
    \item Being able to collaborate on creating and using annotations.
    \item Creating a set of waypoints that trace a path through the data and that can be shared with others.
\end{itemize}

User stories (a more comprehensive list of user story examples for DP 4 is provided in Appendix-Table~\ref{tab_dp_stories}):
\begin{itemize}
    \item As Olivia, I want to keep notes within my threat-hunting tool linking together a set of data, so that I can use the notes to help me construct a mental model and to inform future hunts.
    \item As Jay, I want to have alternative ways to filter data by annotations (metadata), to facilitate reviewing and working on annotations created by myself or other THs during an investigation.
\end{itemize}

\subsection{Integrate External Resources}

Finally, DP 5 \textit{Integrating External Resources} relates to the integration of common external resources into primary threat hunting tools to streamline THs’ processes. 

Related problems include:
(a) context switching caused by internet searches for information supporting hunting activities;
(b) online information resources (e.g., documentation sites) can be difficult to locate or search through;
and (c) the first or most obvious search results may be second-hand information (e.g., from users on Reddit, Stack Overflow), which could be poisoned by bad actors.

Potential solutions could consider:
\begin{itemize}
    \item Automate the lookup of executable hashes and integrate into the TH tool.
    \item Integration of common external bookmarked resources into the TH tool.
\end{itemize}

User stories (examples for DP 5):
\begin{itemize}
    \item As Olivia, I want to have common, external, and previously bookmarked resources integrated into my threat hunting tool so that I can access these resources in a more streamlined way within my threat hunting process and ensure the information I am referencing is accurate.
    \item As Jay, I want to have an automated lookup of executable hashes, so I can know immediately if the executable is custom or known.
\end{itemize}

\section{Discussion}
\label{sec_discussion}

In this section, we discuss the implications of the findings from our study (Subsection~\ref{sec_sub_discussion_implications}). We also reflect on the limitations of our study (Subsection~\ref{sec_sub_discussion_limitations}), and we conclude with future work considerations (Subsection~\ref{sec_sub_discussion_future_work}). 

\subsection{Implications}
\label{sec_sub_discussion_implications}

This section explores the broader implications of our study's findings, highlighting their significance for cybersecurity and threat hunting context. 

\subsubsection{Empirical evidences applicable to other threat hunting contexts}

Our proposed model (presented in Section~\ref{sec_sub_results_model}) is rooted in empirical evidence gathered from observing THs in a real-world setting, distinguishing it from theoretical cybersecurity models presented by other authors (e.g., \cite{andrade_cognitive_2019, gunter2021}). 
Observing THs in a natural setting was a special privilege. It allowed us to gain in-depth insights that were considered to apply to not only these THs but to others (as suggested by cybersecurity experts associated with our industry partner).

The profile of the THs we studied includes individuals working in collaborative team environments and consulting for clients within the private sector using the same tools. 
In addition, THs in our study used the industry partner tool (OpenText, ArcSight Intelligence) as their main tool. 
However, despite our focused context, the findings described in our results apply to other tools used by THs in general (e.g., including Splunk and LogRhythm, which offer SIEM and UEBA capabilities in a single product).

Our findings also resonate closely with established threat hunting models, such as TaHiTI \cite{vanos2018}, affirming their relevance and potential applicability to the broader threat hunting scope. Unlike TaHiTI, which focuses solely on structured hunting, our proposal accommodates both structured and unstructured hunting workflows. This dual faceted approach aligns with other empirical descriptive TH processes (e.g., \cite{maxam2024, badva2024}), however, our model maintains a balanced level of detail to enhances its usability, relevance, clarity, and applicability.

Additionally, the 23 themes that emerged from our thematic analysis presented in Section~\ref{sec_results} and the five design propositions we put forward in Section~\ref{sec_design_propositions}) pave the way for future work. For instance, our reported findings can be informative and used as a baseline for researchers and practitioners to create new data collection instruments (e.g., surveys) to continue exploring in their research settings.
We cover other possible ways to continue in Subection~\ref{sec_sub_discussion_future_work}.

\subsubsection{Threat hunter cognitive efforts}

During our observations, we noticed the significant cognitive effort that THs invest in their memory, underscoring the importance of supporting the construction, externalization, and sharing of a robust mental model.
Considering the clarity of the TH's mental model directly impacts the effectiveness of reporting, communication, and collaboration with clients and peers, we explicitly incorporated the transition of a TH's mental model from \textit{fuzziness} to \textit{clarity} into our model.
This transition highlights the cognitive processes involved, often hidden or underemphasized in other references, even those concentrating on cognitive aspects like Andrade and Yoo \cite{andrade_cognitive_2019}. 
By expanding on the \textit{Mental Model}, we aim to represent the intricate and indirectly observable processes of information processing that are crucial during threat hunting sessions. 

Our model seeks to elucidate these human thought processes without overloading the visual representation with extraneous activities, thus maintaining clarity and focus on core cognitive tasks. Unlike many other models within the security scope that lean towards developing automation and AI, our model emphasizes supporting the human aspects of threat hunting. We aim to capture the intentions and activities performed by THs rather than automating cognitive tasks such as pattern recognition---hence preserving the essence of human cognitive engagement in cybersecurity.

\subsubsection{Bringing attention to new activities}

Our proposed \THmodel~strives to balance practicality and applicability to support the design propositions derived from our findings and our prior experience designing tools. It aligns with the practical, applicable nature of TaHiTI model\cite{vanos2018}, avoiding the overly detailed approach of other references (e.g., \cite{maxam2024, andrade_cognitive_2019, trent2019modelling}) while also steering clear of the high-level abstraction (e.g., \cite{badva2024, gunter2021}). This balanced approach ensures that our proposed model remains accessible and valuable for practitioners, enabling them to apply our insights effectively in real-world threat hunting scenarios. 
Considering this and the specific context of our observations (i.e., THs using SIEM and UEBA tools, working with clients using the same tools), we have emphasized two critical activities in our representation: \textit{Annotate} and \textit{Materialize Mental Model}. These strategic additions underscore the practical necessity for THs to document their findings and thoughts throughout their investigative processes (\textit{Threat Hunting Session}).

\subsection{Limitations}
\label{sec_sub_discussion_limitations}

Although we developed a rigorous study protocol, our study still has a few limitations. Some limitations are inherent to qualitative studies involving observations, such as the effect of participants behaving differently when knowing they are being observed. 
In the following subsections, we list some of the limitations and mitigation actions as well as recommendations for future studies.

\subsubsection{Recruitment process bias} 
Our industry partner managed the recruitment process, as the ``threat hunter'' role is new and uncommon and we had limited access to potential participants. While this approach facilitated access to experienced THs and detailed information that might otherwise be unattainable, it also introduced potential bias since the sample was drawn from a specific industry subset---a limitation we could not mitigate within the scope of our study. Nonetheless, the depth of information gained through this partnership provided valuable insights into the practices and challenges THs face.

\subsubsection{Observations of replayed threat hunting sessions}
During our observation sessions, the TH participants presented how they worked, shared the tools they used and processes followed, and demonstrated their hunting sessions (including actual client data discussing real scenarios, even though not ``live''). The participants selected the content they wanted to share with us as a replay of a previous threat hunting session, with thought and consideration about what they wanted to showcase. 
Consequently, they might have omitted specific steps or presented their hunting activities in a different light than they would have in real time. 
This observational approach provided valuable insights, revealing unspoken or unconscious practices that would have been difficult to uncover through interviews, as participants may not recall or articulate the full nuance of their activities. 
However, we acknowledge that replayed sessions may present a more polished or linear narrative than live investigations, introducing a potential narrative bias. To account for this, we triangulated the observations across multiple participants and sessions. Additionally, our model explicitly emphasizes the iterative and non-linear nature of threat hunting, reflecting how activities such as \textit{Search / Explore}, \textit{Annotate}, and \textit{Materialize Mental Model} can recur and overlap rather than unfold sequentially.

\subsubsection{Researcher bias and interpretive validity} 
Due to the absence of video or audio recordings to verify our observation sessions, we relied on capturing notes that seemed most relevant at the moment. This reliance on the researcher's memory may have led to the omission of some important details. Additionally, some content was intentionally left out of the notes if it was considered to be business-sensitive content. To mitigate researcher bias, we included multiple researchers during the observation sessions (as many as were free at the often unusual times of day) and the data analysis sessions. We also conducted numerous agreement sessions among the research team. Furthermore, we implemented validation processes with the TH participants in our study (member checking) and our industry partner (experts in cybersecurity).

\subsubsection{Population generalizability} 
The findings reported in our study may only be generalizable to THs who use SIEM and UEBA tools, work in team settings, and need to share their findings with clients and other stakeholders. 
While this is a common scenario and we received positive feedback from cybersecurity specialists (industry partner) on the representativeness of our findings, we acknowledge that these results may only apply to some THs due to the diverse range of tools, techniques, and hunting styles they employ. 
Therefore, further research is essential to support the generalizability of our results by conducting new studies in different threat hunting contexts and organizations for a more comprehensive understanding of threat hunting.

\subsection{Future Work}
\label{sec_sub_discussion_future_work}

Looking ahead, we recommend future research exploring both the threat hunting process and the human aspects of the threat hunter. In addition to the future work considerations already discussed (such as validating these findings across different organizations), we highlight additional opportunities for research and potential directions in this section.

\subsubsection{Reports and visualizations}

Future work should investigate the nature of the reports and visualizations created by THs. In addition, explore how the documentation already generated by the THs during the threat hunting session can be streamlined or enhanced.
Another avenue for expanding the scope of visualizations in the TH context is to explore how the findings presented in our study can help address gaps in cybersecurity situational awareness. As identified by Jiang \emph{et al.}~\cite{jiang2022}, future situational awareness visualizations should incorporate information for higher level decision-makers, utilize novel data sources, facilitate collaboration and information sharing, and promote user-centred design.

\subsubsection{Threat hunting cognitive aspects}

Further studies should explore in greater depth how different mental models and cognitive strategies impact the effectiveness of threat hunting. Understanding these impacts could lead to the development of more efficient and effective threat hunting strategies and tools.

\subsubsection{Collaboration aspects}

Another interesting area for future work is exploring TH team dynamics and collaboration aspects such as handover and meetings with clients. 
To support insights, different models and theories could be further explored in this context, such as a theory of team mental models~\cite{Klimoski1994}, transactional memory~\cite{peltokorpi2008} and distributed cognition~\cite{mancuso2020}.

\subsubsection{New tool design}

Our work presents several design considerations for tool developers to explore. For example, the themes and detailed descriptions we provide can serve as a baseline for validating existing needs, uncovering new ones, and supporting the development of new design propositions for use in similar contexts. Since we do not rank or prioritize our design propositions, nor claim they are complete, a logical next step would be to engage with tool designers and product teams to assess their impact within specific contexts and investigate the design propositions further through prototyping.

\subsubsection{The effects of advances in AI}

Researchers such as Mahboubi~\emph{et al.}~\cite{mahboubi_evolving_2024} and Nour~\emph{et al.}~\cite{nour2023_th_enterprise} noted that the adversarial use of generative AI has heightened concerns about the development of compelling social engineering campaigns and the creation of malicious software and resources for more sophisticated attacks. 
A recent publication by Jhanjhi~\cite{jhanjhi_utilizing_2025} advances this discussion. It highlights the need to explore the limitations of traditional threat hunting and the potential of generative AI to enhance behavioural analysis, anomaly detection, and suspicious activity identification. 
Thus, a new area of discussion supported by empirical evidence could focus on how AI-driven tools are reshaping workflows, decision-making, and collaboration, with potential shifts in threat hunting strategies.
Beyond these perspectives, future research could investigate how the mental model of THs may evolve in response to the dynamic nature of threat hunting systems and advancements in AI.

\section{Conclusions}
\label{sec_conclusions}

As an emerging role in cybersecurity, the study of threat hunters remains relatively nascent. Our study represents a step towards offering valuable insights into threat hunters' cognitive processes, workflows, and tooling needs. 
Through comprehensive data analysis of observation sessions, we identified 23 themes, shedding light on the multifaceted role of threat hunters, revealing how they build mental models, and emphasizing the importance of tools that support these cognitive efforts. 
Our findings also underscore the collaborative nature of threat hunting and the need for enhanced communication and information-sharing tools. 
Finally, we presented five design propositions to develop more effective threat hunting processes and tools by addressing cognitive challenges and focusing on human-centric approaches. 
Future work should build on these findings to refine and expand our understanding across diverse threat hunting contexts.

\section*{Acknowledgments}
The authors would like to thank and report the financial support provided for project research during this study by Mitacs Canada, OpenText Corporation, and the Natural Sciences and Engineering Research Council of Canada (NSERC).
The authors also thank all the participants of this study and the reviewers of this manuscript.

\bibliographystyle{IEEEtran}
\bibliography{bibliography}

\section*{Appendix}
\label{appendix}

\begin{figure*}[h]
    \centering
    \includegraphics[width=0.8\linewidth]{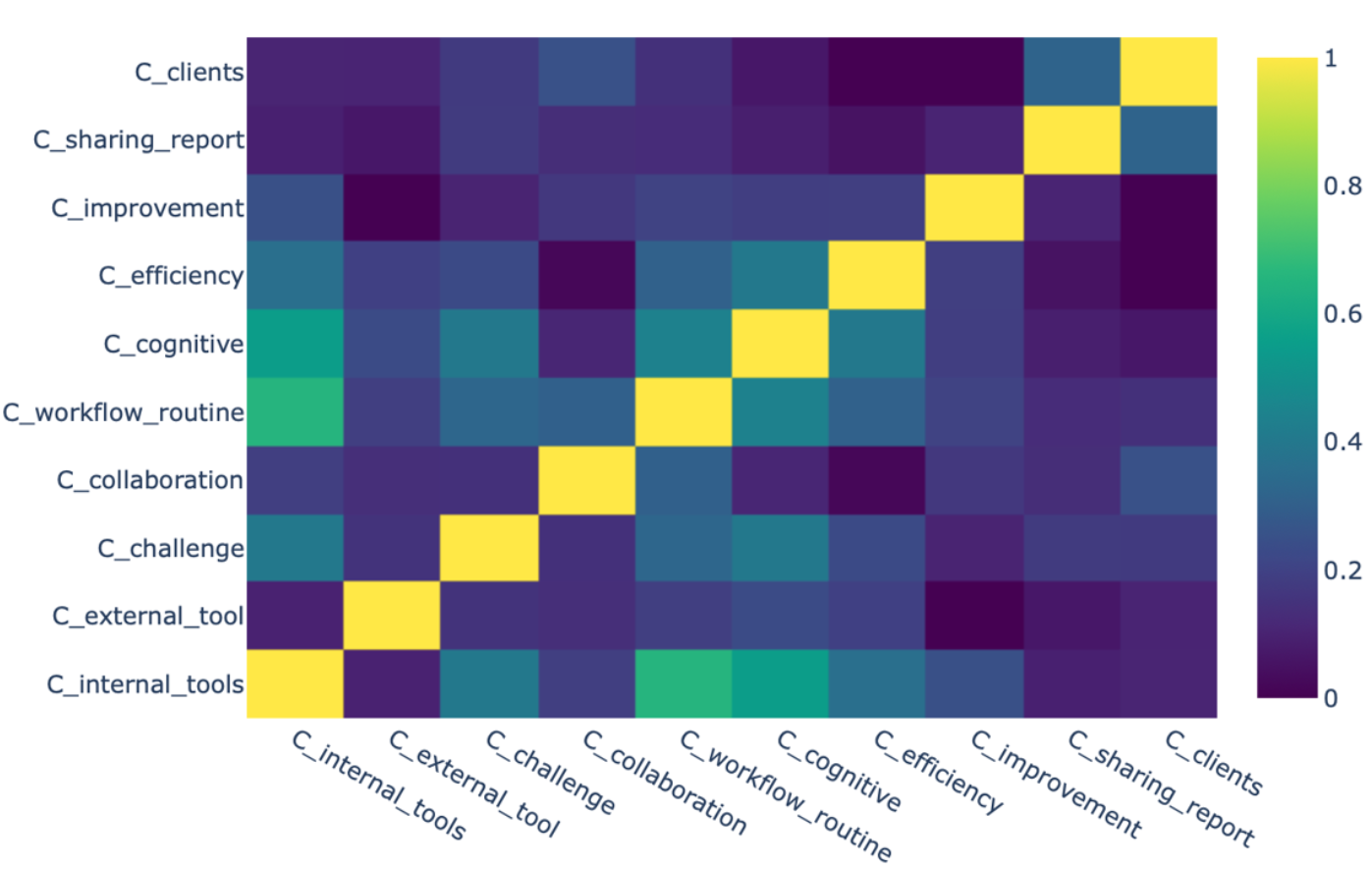}
    \caption{Which categories often or do not often appear on the same cards? Similarity of categories calculated using Jaccard Score. Note: since \textit{Internal\_Tools} was the category most associated to cards (81 of 103), it was expected to see this category with higher scores (close to 1 / Yellow).}
    \label{fig:categories_similarity}
\end{figure*}

We are adding information to support the data analysis process description (coding scheme). 
Table~\ref{tab_categories_description} shows the list of categories descriptions and examples of the cards associated with them. 
Table~\ref{tab_top_cards} shows the three cards with more categories associated with them. 
Figure~\ref{fig:categories_similarity} shows the categories' similarity based on the cards related to them.
Table~\ref{tab_themes_description} shows the list of themes (alphabetically ordered) and their brief description---full theme description including an example of observation note was described in Section~\ref{sec_results}.
Finally, Table~\ref{tab_dp_stories} shows user story examples for each Design Proposition---presented in Section~\ref{sec_design_propositions}.

\begin{table*}[!hb]
    \caption{Categories' Description}
    \label{tab_categories_description}    
    \begin{tabular}{{p{2cm}p{6cm}p{8.5cm}}}
        \toprule
        \textbf{Category Name} & \textbf{Category Description} & \textbf{Example of Card} (Observation Note)\\
        \toprule
         Challenge & Obstacle that a threat hunter faces when attempting to accomplish a goal, for example, due to their tool's implementation or feature set & \textit{A lack of data from the client or lack of visibility into the client's system is a challenge for threat hunters, as it means ``you don't have a map to navigate.''} \tiny{[Card Id 103]} \\
         \midrule         
         Clients & Involves interaction with the client & \textit{The threat hunter will send an email to the client's security team if there is any critical finding to be addressed. Sometimes, the noted activity is due to penetration testing.} \tiny{[Card Id 37]} \\
         \midrule         
         Cognitive & Places a significant cognitive burden on threat hunters & \textit{A threat hunter remarked that it is ``almost impossible to memorize these codes'' when examining the details of an event, before using an internet browser to search for the given code.} \tiny{[Card Id 58]}\\
         \midrule
         Collaboration & Collaboration activities with internal or external individuals & \textit{Findings and logs from each hunt are shared with the next threat hunter through OneNote.} \tiny{[Card Id 55]} \\
         \midrule
         Efficiency & Time-consuming task that a threat hunter performs relatively often and has the potential to be automated & \textit{Threat hunters were observed to frequently copy and paste. Sometimes, using Notepad as a temporary file. Not necessarily to external tools. Sometimes, just saving and restoring settings (e.g., filters) within their threat hunting tool.} \tiny{[Card Id 23]} \\
         \midrule                           
         External\_Tool & Any and all tools aside from the threat hunting tool developed internally, including digital notebooks, internet browsers, and other external resources & \textit{Threat hunters frequently change context from their threat hunting tool to search elsewhere on the internet or copy info to/from OneNote.} \tiny{[Card Id 65]} \\
         \midrule         
         Improvement & Improvement, feature, or solution proposed by the research team or participants & \textit{Threat hunters feel pressured when deciding when to stop hunting. It may help to include features (e.g., timer, percentage coverage, checklists) that support the threat hunter in deciding when to conclude the hunt.} \tiny{[Card Id 66]} \\
         \midrule         
         Internal\_Tool & Pertains to the internal threat hunting tool & \textit{Threat hunter created a new browser tab for each relevant entity related to the current entity (breadth-first search approach).} \tiny{[Card Id 20]} \\
         \midrule         
         Sharing\_Report & Involved in the creation and sharing of reports with stakeholders & \textit{The threat hunter must create a ``big picture'' of the attack, for themselves, other threat hunters, management, and others.} \tiny{[Card Id 68]} \\
         \midrule         
        Workflow/Routine & Describes the order in which a threat hunter completes tasks, including guidelines and other practices that a threat hunter incorporates into their process & \textit{Threat hunter kept a list of compromised machines that needed to be quarantined as they could otherwise be used as launch points for further attack.} \tiny{[Card Id 21]} \\
        \bottomrule
    \end{tabular}
\end{table*}

\clearpage
\begin{table*}[]
    \centering
    \caption{Cards with more categories associated with (total of seven each) and the final theme related to them}
    \label{tab_top_cards}    
    \begin{tabular}{{p{8.5cm}p{5cm}p{3cm}}}
        \toprule
        \textbf{Card} (Observation Note) & \textbf{Categories Associated} & \textbf{Theme Associated} \\
        \toprule
        \textit{Behavioral analysis by the threat hunter during a hunt requires pre-existing communication with client to understand acceptable/expected patterns of behavior.} \tiny{[Card Id 41]}  & Challenge; Clients; Cognitive; Collaboration; External\_Tool; Internal\_Tool; Workflow/Routine & Client Collaboration and Dependent Processes \tiny{(\ref{sec_sub_themes_2_2})}  \\
        \midrule        
        \textit{Threat hunters would like a mind map integrated into \THtool~ for tracking progress and suspicious events during a hunt. This feature was noted as potentially already in progress.} \tiny{[Card Id 74]} & Challenge; Cognitive; Efficiency; Improvement; Internal\_Tool; Sharing\_Report; Workflow/Routine & Mental Model of Active Threat Hunt Activity \tiny{(\ref{sec_sub_themes_1_9})}   \\
        \midrule        
        \textit{Idea: enable the workflow in \THtool~ to focus around capturing the significant events of an attack or hunt. Importantly, this is related to how a threat hunter builds a mental model and would help to capture the sequence of events. The information captured through this interaction with the tool could be shared to aid collaboration and reporting.} \tiny{[Card Id 80]} & Challenge; Cognitive; Collaboration; Improvement; Internal\_Tool; Sharing\_Report; Workflow/Routine  & Mental Model of Active Threat Hunt Activity \tiny{(\ref{sec_sub_themes_1_9})} \\        
        \bottomrule
    \end{tabular}
\end{table*}

\begin{table*}[htbp]
    \centering
    \footnotesize
    \caption{Themes' Description}
    \label{tab_themes_description}    
    \begin{tabular}{{p{3.5cm}p{13cm}}}
        \toprule
        \textbf{Theme Name} & \textbf{Theme Description} \\
        \midrule
        Active Threat Hunting Process & Workflow or routine related to the active threat hunting process (e.g., steps and checklists). It is also associated with the structured, standardized, and objectives for a particular TH team, such as using heuristics, guidelines, or checklists. 
        \\
        \midrule
        Attacker Strategy & Describes the threat hunter's process for tracing the activities and movement of an attacker through the system and learning the patterns of the attacker to find related events or activities. 
        \\
        \midrule
        Client Collaboration and Dependent Processes &	Interactions with the client and processes that are unique or dependent on the client for whom the process is being conducted. 
        \\
        \midrule
        Collaboration with Threat Hunting Tool Maintainers and Developers & Processes through which a threat hunter can effect change in their tool through the tool's maintainers and developers. 
        \\
        \midrule
        Collaborative Active Threat Hunting	& Active threat hunting, performed with real-time communication and collaboration with other threat hunters looking at the same data. 
        \\
        \midrule
        Data Availability Limitations &	Problems with missing data, and not in how the data is being processed or analyzed, which is strictly a failure of the tool performing the analysis. 
        \\
        \midrule
        Documentation of Active Threat Hunting Findings	& The information recorded during threat hunting activities and how that information is created, represented, applied, and shared. 
        \\
        \midrule
        Ease of Pivoting and Exploring in UIs & UI-specific challenges, ideas, and improvements related to how the threat hunter pivots on events. It is purely related to presentation and user interaction. It is connected to ``Missing Pivot Points Between Correlated Events or Groups of Events'' theme, which describes data processing rather than presentation. 
        \\
        \midrule       
        Event Search Capabilities & (tooling) capabilities to support threat hunters searching for events (e.g., using filters or keywords). 
        \\
        \midrule
        Feedback Loop Between a TH and Their Tool During an Active Threat Hunt & Threat hunter's ability to inform the tool of their findings and next steps. It is also related to the tool's ability to support the threat hunter using this additional context (e.g., by filtering, suggesting or highlighting information and views). 
        \\
        \midrule
        Frequent use of Memory & Using the threat hunter's memory to store helpful information, necessary or relevant for active threat hunting purposes, such as process names and status codes. 
        \\
        \midrule
        Handover Process & Protocol and resources THs use to hand over information across shifts (or during shift changes). 
        \\
        \midrule
        Information Resource Challenges	& Challenges associated with using information resources such as websites and documentation. It applies to static information sources and is not associated with compute resources. 
        \\
        \midrule
        Internal to External Data Linking & Linking internal threat hunting data to external resources, for example, linking Windows process names in event logs to their place in the online documentation. 
        \\
        \midrule
        Internal Tooling Capabilities, Challenges, and Opportunities & Limitations, inefficiencies, ideas, and strong points in the threat hunter's current tools (not resources). It only applies to cards specific to the internal tool (design/UI) that are also not strongly related to other themes. This theme acts as a catch-all for cards related to the internal tool without a specific theme. For cards that include tooling challenges but also relate to other themes, the ``internal tools'' category is applied instead, and an explicit edge to this theme is omitted. 
        \\
        \midrule
        Limitations of UEBA	& Limitations inherent to anomaly detection approach (attackers can create noisy activity to reduce the likelihood of any of their malicious activity being flagged as ``anomalous''). 
        \\
        \midrule    
        Mental Model of Active Threat Hunt Activity	& Internal (in individual’s head) or external (in software, on paper, etc.) organization or conceptual model the TH builds of notable or suspicious events and their story (or timeline).  
        \\
        \midrule
        Mental Model of Client's System & Internal (in the individual's head) or external (in software, on paper, etc.) organization or conceptual model the TH builds of the client's environment. This mental model contextualizes the TH's activities and understanding and provides the TH with their bearings during the hunt and intuitions. 
        \\
        \midrule
        Missing Pivot Points Between Correlated Events or Groups of Events & Missing ability to pivot on events to detect anomalies across related entities (e.g., navigate to similar events or entities instead of individual entities). It is distinct from the UI issues as it is purely the quality/existence of these associations after processing by the backend.  
        \\
        \midrule
        Reporting & Techniques, tools, and processes used (by anyone, e.g., THs or clients) to generate and communicate reports on threat hunting activities, such as findings and results. 
        \\
        \midrule
        Significant Event Marking & How and why events are annotated. For instance, ``how'' can be the tool tags, such as the briefcase icon, and ``why,'' the bookmarks and communication with the client. 
        \\
        \midrule
        Technical Skills and Experience	& Threat hunters' technical skills and knowledge background, including operating systems, system administration, computer networks, and other areas. 
        \\
        \midrule
        When to Stop Hunting? & Heuristics or frameworks used by THs to decide when to conclude the active threat hunt. 
        \\        
        \bottomrule
    \end{tabular}
\end{table*}

\begin{table*}[]
    \centering
    \caption{Examples of user stories for each design proposition}
    \label{tab_dp_stories}    
    \begin{tabular}{{p{3cm}p{13.5cm}}}
        \toprule
        \textbf{Design Proposition}  & \textbf{User Story} \\
        \toprule
        \vspace{1pt} DP1. Creating a Story or Timeline of Events & 
        \begin{itemize}
            \item As Olivia, I want to externalize (draw) the story of the hunt I am working on so that I can clarify my thoughts and reduce my cognitive load.
            \item As Thomas, I want to add notes to my externalized mental model so that I can explain the data to myself and others.
            \item As Olivia, I want to share my externalized mental model with other THs to align our mental models and share our findings during an active hunt.
            \item As Jay, I want to have a way to save interesting items and sort items into abstract containers so I can review the items easily and begin to clarify my thoughts.
            \item As Thomas, I want to add events, machines, and users directly from my hunting tool to my externalized mental model so that my externalization has backlinks to the data and includes relevant context.
            \item As Thomas, I want to capture the temporal dimension of my externalized mental model so that I can describe the influence and spread of suspicious activity.
            \item As Olivia, I want to use my externalized mental model as a starting point to generate (elucidate) formal reports for clients and management.
            \item As Jay, I want my externalized mental model to be used by my hunting tool to assist in exploration (e.g., suggestions, in context analytics, or shortcut for searching).
            \item As Olivia, I want my threat hunting tool to suggest whether or not to stop hunting based on the findings and comprehensiveness of my externalized mental model so that I can conclude hunts with confidence and avoid ending too early (potentially missing evidence), or ending too late (and taking time away from other hunts).
            \item As Jay, I want to be able to save and use a previous workflow for a new hunt.
            \item As Thomas, when I start my hunt I want to be able to examine the stories from the previous shift to be able to orient myself in the current threat situation.
        \end{itemize}
                \\
        \midrule        
        \vspace{1pt} DP2. Visualizing / Navigating Connections (Spatial) & 
        \begin{itemize}
            \item As Thomas, I want to visualize and see where I am within a spatial map of the client's system/network so that I can orient more easily to understand the activity I'm seeing.
            \item As Jay, I want to navigate from one entity to other related entities based on interactions in the data, so that I can explore the connections and make sense of what's happening.
            \item As Olivia, I want to explore the event details of several different events without losing track of where I am, so that I can synthesize and make sense of what's happening.
            \item As Thomas, I want to see what happened before and after an event, so that I can identify correlations of events related in time.
        \end{itemize}
            \\
        \midrule        
        \vspace{1pt} DP3. Creatively Expanded Search & 
        \begin{itemize}
            \item As Jay, I want to search for patterns that are similar to known attack strategies and be able to ``find similar'' patterns to one I'm seeing.
            \item As Thomas, I want to search for patterns in the data that are similar to observed patterns in the attacker's activities so far, so that I can discover the full scope of compromised machines.
        \end{itemize}        
            \\        
        \midrule        
        \vspace{1pt} DP4. Waypoints and Note-Taking & 
        \begin{itemize}
            \item As Olivia, I want to keep notes within my threat-hunting tool linking together a set of data, so that I can use the notes to help me construct a mental model and to inform future hunts.
            \item As Jay, I want to have alternative ways to filter data by annotations (metadata), to facilitate reviewing and working on annotations created by myself or other THs during an investigation.
            \item As Thomas, I want to annotate the event data in my hunting tool with waypoints, so that I can record my path and so that others and I can retrace my steps and ensure the full coverage of the data.
            \item As Olivia, I want an integrated view and history of annotations so that I can collaborate more easily with my team and discuss team agreement of the notes.
            \item As Olivia, I want to enhance the annotations with a agreement/disagreement feature so that I can collaborate more easily with my team and coordinate different beliefs across the team.
        \end{itemize}
                \\  
        \midrule        
        \vspace{1pt} DP5. Integrating External Resources & 
        \begin{itemize}
            \item As Olivia, I want to have common, external, and previously bookmarked resources integrated into my threat hunting tool so that I can access these resources in a more streamlined way within my threat hunting process and ensure the information I am referencing is accurate.
            \item As Jay, I want to have an automated lookup of executable hashes, so I can know immediately if the executable is custom or known.
        \end{itemize}        
                \\           
        \bottomrule
    \end{tabular}
\end{table*}

\end{document}